\documentclass[%
 aps,
 prx,
 amsmath,amssymb,
 reprint,%
]{revtex4-2}

\usepackage{graphicx}
\usepackage{dcolumn}
\usepackage{bm}
\usepackage{mathptmx}
\usepackage{tabularx}
\usepackage{graphicx}
\usepackage{multirow}
\usepackage{siunitx}
\usepackage{booktabs}
\usepackage{xcolor}
\usepackage[colorlinks=true,citecolor=blue
]{hyperref}
\usepackage{natbib}

\begin{document}
	
	\title{Modeling of electric double layer at solid-liquid interface with spatial complexity}
	
	\author{Cherq Chua}
	\affiliation{Science, Mathematics and Technology, Singapore University of Technology and Design, 8 Somapah Road, Singapore 487372}
	
	\author{Chun Yun Kee}
	\affiliation{Science, Mathematics and Technology, Singapore University of Technology and Design, 8 Somapah Road, Singapore 487372}
	
	\author{L. K. Ang}
	\email{Corresponding Author. ricky\_ang@sutd.edu.sg}	
	\affiliation{Science, Mathematics and Technology, Singapore University of Technology and Design, 8 Somapah Road, Singapore 487372}	

	\author{Yee Sin Ang}
	\email{Corresponding Author. yeesin\_ang@sutd.edu.sg}
	\affiliation{Science, Mathematics and Technology, Singapore University of Technology and Design, 8 Somapah Road, Singapore 487372}	

\begin{abstract}
Electrical double layer (EDL) is formed when an electrode is in contact with an electrolyte solution, and is widely used in biophysics, electrochemistry, polymer solution and energy storage.
Poisson-Boltzmann (PB) coupled equations provides the foundational framework for modeling electrical potential and charge distribution at EDL.
In this work, based on fractional calculus, we reformulate the PB equations (with and without steric effects) by introducing a phenomenal parameter $D$ (with a value between 0 and 1) to account for the spatial complexity due to impurities in EDL. 
The electrical potential and ion charge distribution for different $D$ are investigated.
At $D$ = 1, the model recover the classical findings of ideal EDL.
The electrical potential decays slowly at $D <$1, thus suggesting a wider region of saturated layer under fixed surface potential in the presence of spatial complexity.
The fractional-space generalized model developed here provides a useful tool to account for spatial complexity effects which are not captured in the classic full-dimensional models.

\end{abstract}

\maketitle	

\textcolor{blue}{\textbf{Introduction. }}Electrical double layer (EDL) is a structure formed when a charged surface is in contact with an electrolyte solution, which has significant applications in the areas of bio-physics \cite{bio, Blank1986ElectricalDL} and polymer science \cite{DC9755900242, NOHARA2003749}. 
In recent years, many technologically important devices making use of the unique functional of EDL have been demonstrated.
One of the examples is the electric double layer capacitor (EDLC). EDLC has high power density with fast charge-discharge speed, making it a good alternative to traditional battery for novel energy storage devices \cite{EDLC}. 
For electric double layer based transistor (EDLT), the EDL is formed at the semiconductor/electrolyte interface, and it exhibits large specific capacitance, which allows lower voltage required as compared to the traditional field-effect transistor (FET) \cite{EDLT}. 
With such advantages, EDLT has been used in synaptic device and neuromorphic system \cite{C8TC00530C, C8NR07133K, doi:10.1021/acsami.8b07234,gao2021artificial}.

Many models have been developed to explain the electrical characteristics of EDL, such as electrical potential and distributions of charge density of the electrolyte solution.
A widely employed model is the Gouy-Chapman model \cite{gouy,chapman}, which is based on the the standard Poisson-Boltzmann (PB) coupled equations \cite{gray2018nonlinear}. 
Although the classical PB equation could provide relatively well approximation to EDL, many experimental results have found significant deviation from the theoretical results obtained with the Gouy-Chapman model \cite{cuvillier,CUVILLIER199819,PhysRevX.6.011007}. 
The deviations are mainly due to some of the simplifications employed in the model.
For example, the standard PB equation has neglected the finite size of the ions, which may overestimate the surface ion concentration at high surface charge density regime \cite{butt2013physics}. 
The inevitable presence of impurities could also serve as extra complexity, which will affect the spatial distribution of the ions and the electrostatic potential \cite{doi:10.1021/acscatal.9b04229,doi:10.1021/acs.jpcc.7b04869}.
While various works have been developed to account for the ion-ion correlation and the finite ion size \cite{Borukhov,BORUKHOV2000221,article,doi:10.1021/j100838a019,doi:10.1063/1.450231, li2019analysis}, how spatial complexity or inhomogeneity at the solid-liquid interface region, which may arise due to the inevitable presence of impurities or electrode roughness, can affect the electrostatic characteristics of EDL remains an open question thus far.

Recently, the applications of fractional modelling provides an effective tool in studying complex physical systems with spatial anisotropy or memory effect \cite{4368612,tarasov}. 
Such complexity effect is substituted by an effective system with fractional dimension, and the corresponding fractional models can be solved using several different mathematical operators developed in fractional calculus.
Many applications of such fractional models have been successful in various areas of fundamental physics, such as quantum field theory \cite{Palmer_2004}, nonlinear dynamics \cite{PhysRevLett.91.034101}, statistical mechanics \cite{paradisi}, thermodynamics \cite{TARASOV2016427} and electrodynamics \cite{nasro,book}.
For applied physics, it has also effectively generalized the Schrodinger equation to account for impurities in solids \cite{PhysRevB.43.2063}, excitons binding energy in 2D materials \cite{excitons-2D}, heat conduction \cite{doi:10.1063/1.4998236}, carrier conduction in organic semiconductor \cite{8388713}, electron field emission \cite{fractionalFN}, space-charge-limited current \cite{doi:10.1063/1.4958944}, light absorption in rough surface \cite{doi:10.1063/1.5039811}, electromagnetic wave propagation \cite{zubair2011exact, Ehsan2020SpaceFractionalBB}, metamaterial \cite{naqvi2010waves} and electromagnetic cloaking \cite{nisar2018cloaking}.
Beyond physical science, fractional modeling are also used broadly in complexity science, such as the modeling of COVID-19 pandemic \cite{tuan2020mathematical, Rajagopal2020AFM}, climate change \cite{Climate}, economics theory \cite{tarasova2016elasticity,math7060509, Tarasov2016memory}, and human-human relationship \cite{KUMAR2021111091}. 

In this paper, we develop a \emph{fractional model} of EDL at an electrode/electrolyte solution interface. 
By generalizing the modified Poisson-Boltzmann (PB) model developed by Borukhov, Andelman and Orland (BAO) \cite{Borukhov} to a fractional dimension $D$ (see below) based on the Stillinger's fractional Laplacian operator \cite{stillinger}, we calculate the spatial profiles of the electrical potential and ion charge distribution with different values of $D$.
Here, the BAO model considers the effect of finite ion sizes in addition to the electrostatic interaction, which is commonly employed to model the potential distribution in EDL consisting of large multivalent ions near to a strongly charged surface \cite{Borukhov}.
The BAO model has been successful in qualitatively explaining the saturation of surface ion concentration observed in experiments \cite{CUVILLIER199819,cuvillier}.
In our proposed fractional model, the phenomenal parameter $D$ (between 0 and 1) is used to mimic the spatial complexity effect, which could arise from electrode roughness, or impurities in the electrolyte solution, which is not captured in the standard PB and BAO models. 
The fractional model recovers the original PB and BAO models at $D$ = 1.

Using our model, we calculate the electrical potential, relation between surface charge density and counter ion concentration at different $D$, which concretely reveals the effect of $D$ on the EDL characteristics, such as the ions concentration profile and the suturation surface charge density.
Beyond the applications of EDL, this fractional model should be relevant to other applications based on PB equations.
The fractional model developed here can be employed to fit the experimental results with $D$ as a fitting parameter, thus offering a fractional modelling route to extend the validity of PB and BAO models for EDL with spatial complexities. 



\textcolor{blue}{\textbf{Model.}} For a charged planar surface in contact with an electrolyte solution consisting of large multivalent counter ions, the standard PB equation is no longer valid in describing the potential distribution since it neglects the non-bonding interaction between the ions. 
This interaction, known as steric effect has been studied in a BAO model \cite{Borukhov}. 
For an asymmetric $1:z$ electrolyte, which has negative multivalent ions of charge $-ze$ and positive monovalent ions of charge $e$, the electrical potential distribution in the solution is
	\begin{equation}\label{eq:modified_PB_1z}
		\nabla^2\psi = \frac{zec_0 }{\epsilon\epsilon_0} \frac{e^{z\beta e\psi}-e^{-\beta e\psi}}{1-\phi_0+\phi_0(e^{z\beta e\psi}+ze^{-\beta e\psi})/(z+1)},
	\end{equation}
where $\psi$ is the electrical potential, $\beta=1/k_BT$ is the thermal energy, $c_0$ is the bulk concentration of the electrolyte and $\phi_0=(z+1)a^3c_0$ is the total bulk volume fraction
of the positive and negative ions, and $a$ is the finite ion size of the electrolytes (assuming both types of ions have the same size).
For a symmetric $z:z$ electrolyte, we have
	\begin{equation}\label{eq:modified_PB_zz}
		\nabla^2\psi = \frac{2 zec_0 }{\epsilon\epsilon_0} \frac{\sinh(z\beta e\psi)}{1-\phi_0+\phi_0\cosh(z\beta e\psi)},
	\end{equation}
where $\phi_0=2a^3c_0$. For both cases, the distribution of the negative ion concentration $c^-$ is related to $\psi$ by  
\begin{equation}
c^-(x) = \frac{1}{a^3}\frac{1}{1+(z+1)\frac{1-\phi_0}{\phi_0}e^{-z\beta{e}\psi}}.
\end{equation}
To generalize the above mentioned models into fractional model, we use the 
Stillinger's second-order fractional Laplacian operator \cite{stillinger}, given by
%
%
	\begin{equation}\label{Stillinger_1D}
		\nabla^{2D}=\frac{d^{2D}}{dx^{2D}}=\frac{d^2}{dx^2}+\frac{D-1}{x}\frac{d}{dx},
	\end{equation}
where $0<D\leq1$ is a fractional parameter.  
Using the asymmetric case as an example, by substituting the non-integer Laplacian operator into Eq. (1), we obtain a fractional BAO model in the form of
	\begin{equation}\label{eq:frac_modified_PB_1z}
		\frac{d^2{\psi}}{d{x}^2}+\frac{D-1}{{x}}\frac{d{\psi}}{d{x}}=\frac{\rho_e}{\epsilon\epsilon_0},
	\end{equation}
and
\begin{equation}
    \rho_e = \frac{zec_0 (e^{z\beta e\psi}-e^{-\beta e\psi})}{1-\phi_0+\phi_0(e^{z\beta e\psi}+ze^{-\beta e\psi})/(z+1)}.
\end{equation}
At $D$ = 1, it recovers the original BAO model in Eq. (1).

\begin{figure} []
    \centering
    \includegraphics[width=8cm]{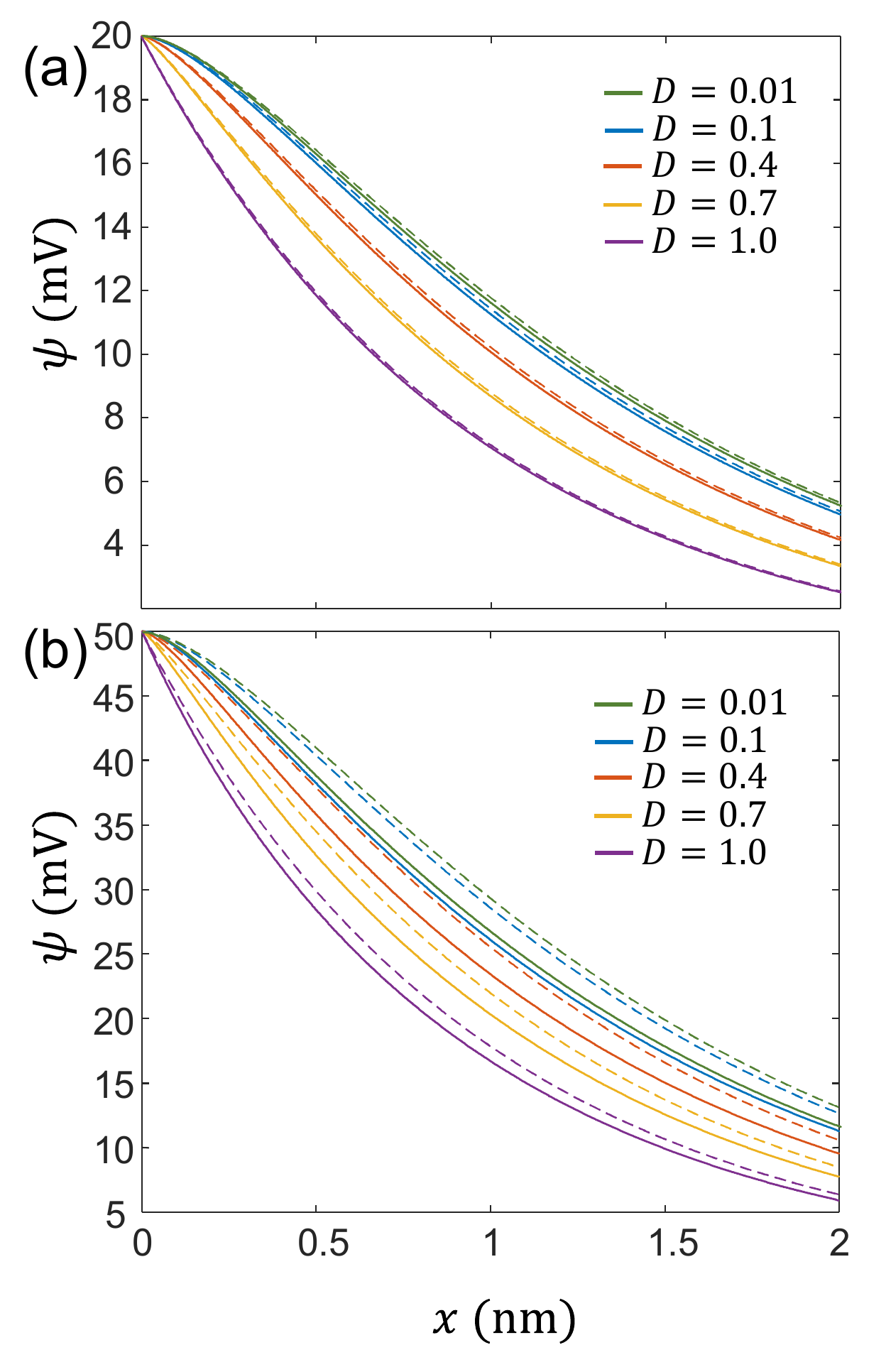}
    \caption{Comparison between full numerical solution (solid lines) and analytical solution of the linearized PB equation (dashed lines) with different $D$ at a) low surface potential ($\psi_0=20$ mV) and (b) high surface potential ($\psi_0=50$ mV). The electrolyte solution is at room temperature with a salt concentration of $c_0=0.1$ M.}
    \label{fig:Fig1}
\end{figure}
\begin{figure}[]
    \centering
    \includegraphics[width=8.5cm]{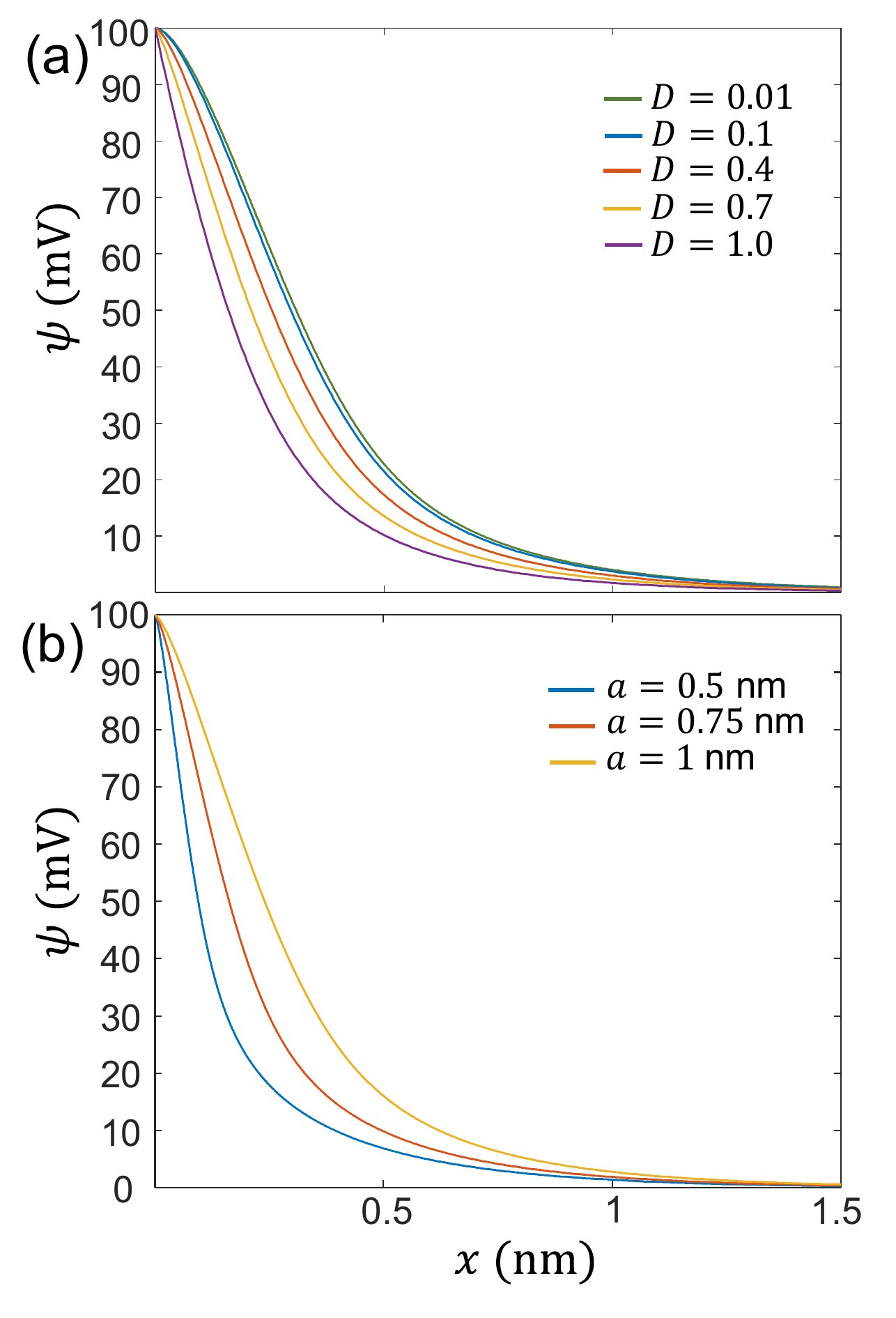}
    \caption{Electrostatic potential profiles at (a) different $D$ with $a=1$ nm, and (b) different ion sizes $a$ with $D=0.5$. The electrolyte solution is at room temperature with $c_0=0.1$ M, $\psi_0=100$ mV and and has an asymmetry electrolyte of $z=4$.}
    \label{fig:Fig2}
\end{figure}
\begin{figure}[]
    \centering
    \includegraphics[width=8.5cm]{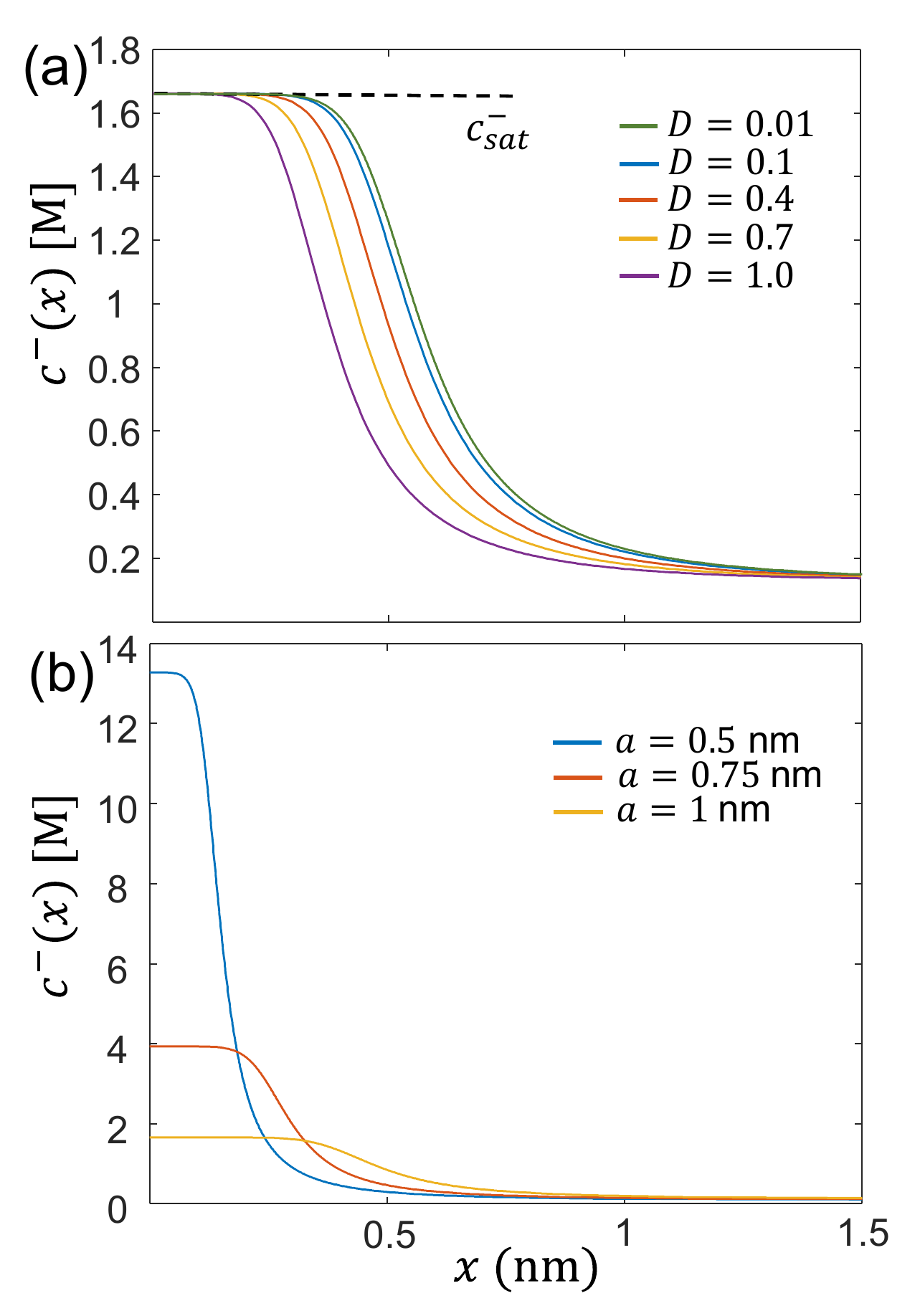}
    \caption{Concentration profiles of negative ions $c^-(x)$ at (a) different $D$ with $a=1$ nm, and (b) different ion sizes $a$ with $D=0.5$. The black dashed line indicates the saturated concentration $c_{sat}^-$. The electrolyte solution is at room temperature with $c_0=0.1$ M, $\psi_0=100$ mV and and has an asymmetry electrolyte of $z=4$.}
    \label{fig:Fig3}
\end{figure}
\begin{figure}[]
    \centering
    \includegraphics[width=8.5cm]{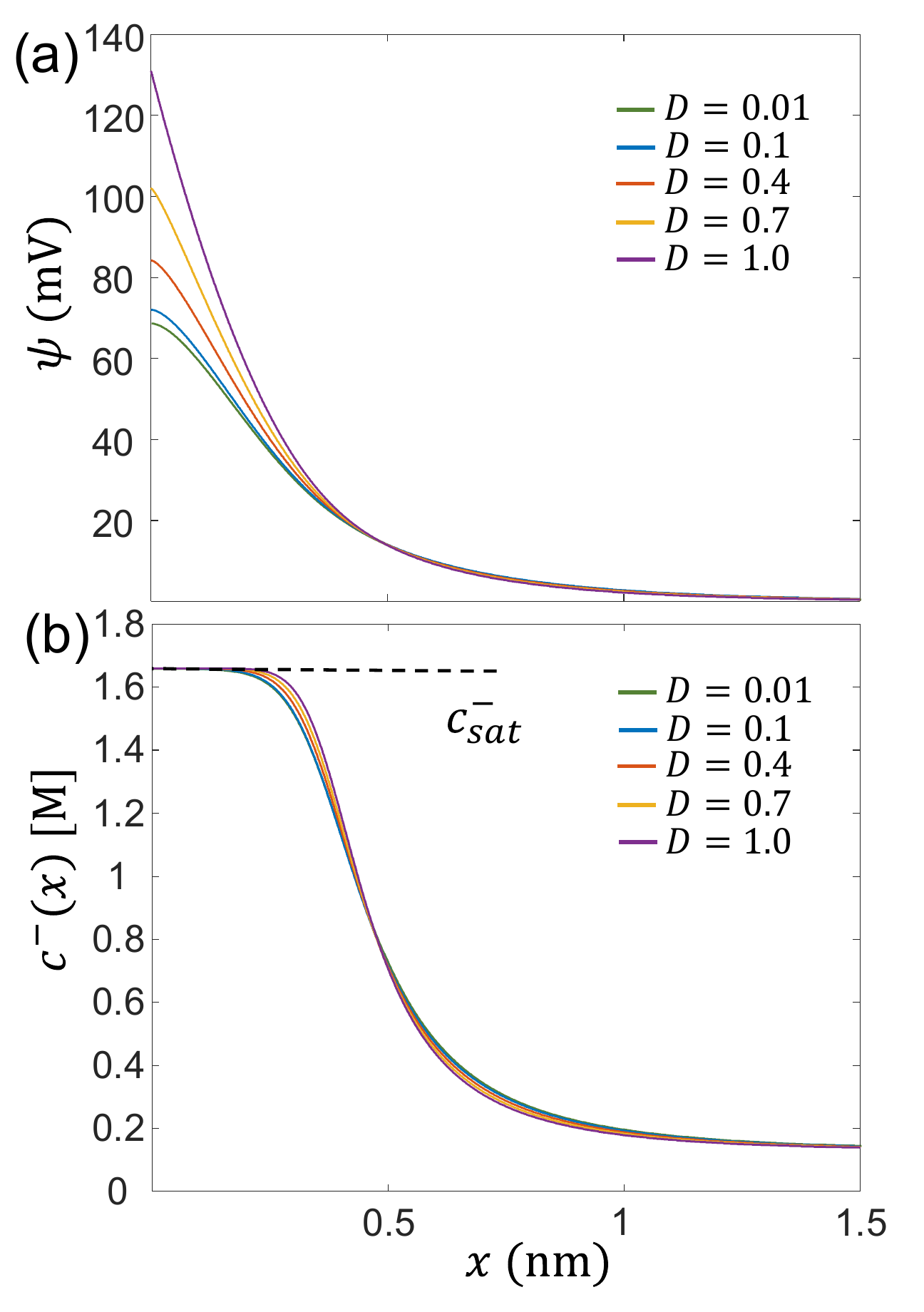}
    \caption{Distribution profiles of (a) electrical potential and (b) negative ion concentration with fixed surface charge density of $\sigma/e=2$ nm$^{-2}$ at different $D$. The black dashed lines indicate the saturated concentration $c_{sat}^-$. The electrolyte solution is at room temperature with $a=1$ nm, $c_0=0.1$ M and has an asymmetry electrolyte of $z=4$.}
    \label{fig:Fig4}
\end{figure}
%
%
%

%
%

%
\begin{figure*} [hbt!]
    \centering
    \includegraphics[width=15.5cm]{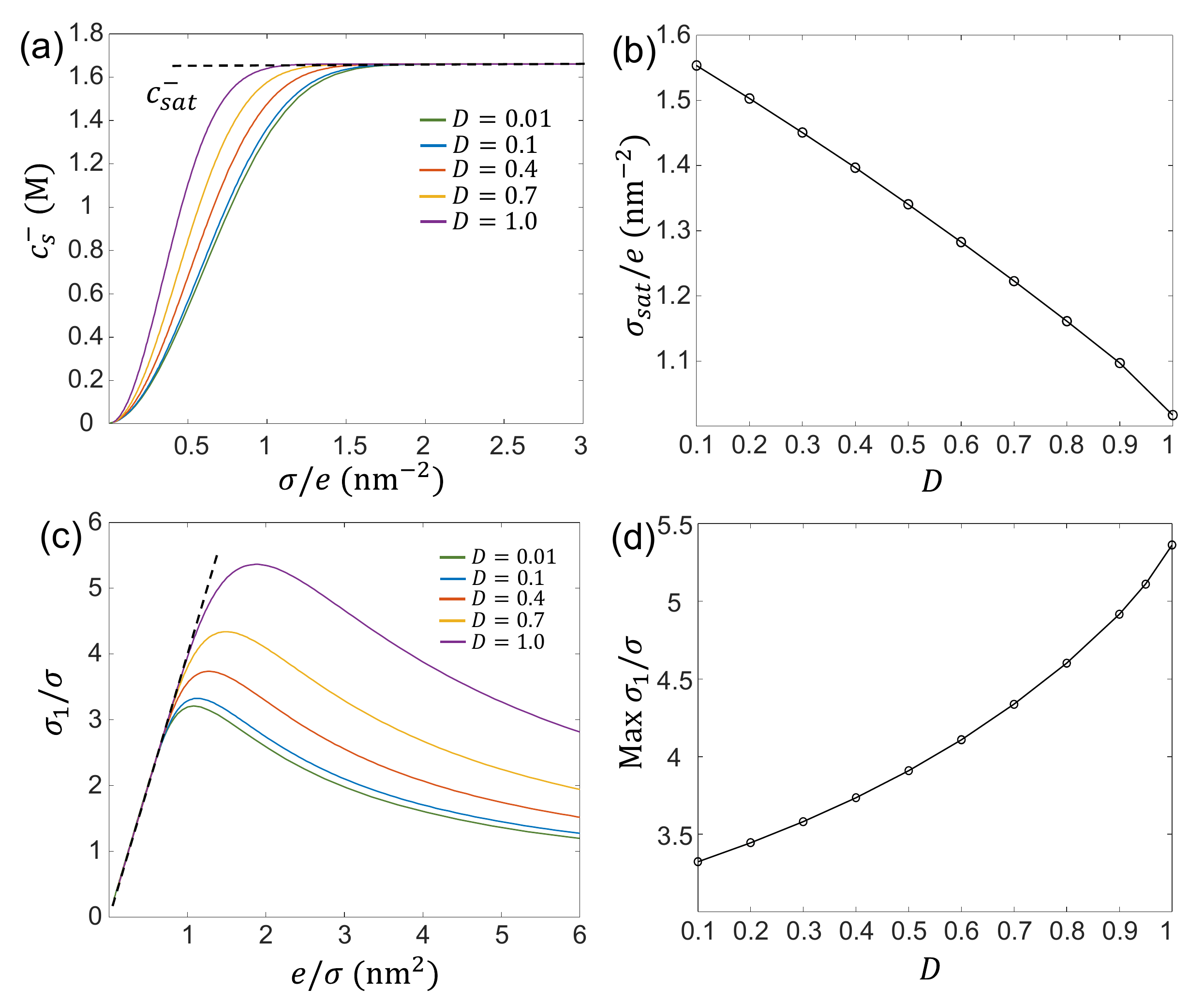}
    \caption{(a) Surface negative ion concentration $c_s^-$ plotted against surface charge density $\sigma/e$; (b) the lowest surface charge density required to reach saturation, $\sigma_{sat}$ plotted against $D$; (c) Ratio of first layer charge density to surface charge density, $\sigma_1/\sigma$ plotted against area per unit surface charge, $e/\sigma$, and (d) the peak value of $\sigma_1/\sigma$ against $D$. The eletrolyte solution is at room temperature with $c_0=1$ mM, $a=1$ nm and and has an asymmetry electrolyte of $z=4$.}
    \label{fig:Fig5}
\end{figure*}
In the standard PB and BAO models, the surface potential $\psi_0$ is linked to the surface charge density $\sigma$ through the Grahame equation \cite{butt2013physics}. 
This relation is, however, no longer valid in the fractional case.
From the electro-neutrality condition \cite{butt2013physics}, the surface charge density $\sigma$ is determined by
\begin{equation}
 \sigma = -\int_{0}^{\infty} \rho_e \,dx.
\end{equation}
%
Combining Eq. (5) and Eq. (7), we obtain a modified Grahame equation, which gives
\begin{equation}
 \sigma = \epsilon\epsilon_0[-\psi^\prime(0)+\left(D-1\right)\int_{0}^{\infty} \frac{\psi^\prime(x)}{x} \,dx].
\end{equation}
Compared to the original Grahame equation with $\sigma = -\epsilon\epsilon_0\psi^\prime(x)$, Eq. (8) consists of an extra integral term, which can be considered as a correction term due to the spatial complexity (e.g. impurities) that is not included in the standard PB and BAO models.



In the case of $a \to 0$ and $z=1$, the BAO model reduces to the standard PB equation, and our fractional model is rewritten as:
    	\begin{equation}\label{eq:frac_stand_PB}
		\frac{d^2{\psi}}{d{x}^2}+\frac{D-1}{{x}}\frac{d{\psi}}{d{x}}=\frac{c_0e}{\epsilon\epsilon_0}\left(e^{\beta e\psi}-e^{-\beta e\psi}\right).
	\end{equation}
At low potential regime: $|{\beta}e\psi|\ll1$, the exponential term in Eq. (9) is expanded for low order only, and the differential equation can be simplified into a linear form, which gives
    	\begin{equation}\label{eq:frac_stand_PB_low}
		\frac{d^2{\psi}}{d{x}^2}+\frac{D-1}{{x}}\frac{d{\psi}}{d{x}}=\frac{2c_0e^2\beta}{\epsilon\epsilon_0}\psi.
	\end{equation}
With the boundary conditions of: (i) a fixed surface potential $\psi(x\to 0) = \psi_0$, and (ii) zero potential at distance far away from the charged surface $\psi(x\to\infty)=0$, an \textit{analytical} solution for Eq. (10) is obtained:
\begin{equation}
 \psi = -2^{\frac{2+D}{2}}\psi_0\left(\kappa x\right)^{\frac{2-D}{2}}\frac{K_{\frac{2-D}{2}}\left(\kappa x\right)}{\Gamma\left(-\frac{D}{2}\right)}. 
\end{equation}
Here, $\kappa =\sqrt{(2c_0e^2\beta/\epsilon_0\epsilon)}$, $K_\alpha(z)$ is the modified Bessel function of the second kind and $\Gamma(z)$ is the gamma function. For $D=1$, Eq. (11) recovers to the analytical solution of the conventional linearized PB model 
\begin{equation}
 \psi = \psi_0 e^{-\kappa x}.   
\end{equation}

\textcolor{blue}{\textbf{Numerical Results and Discussions. }} The generalized fractional PB model with steric effect is solved with the following numerical scheme. The potential distribution profile is obtained by solving Eq. (5) with two boundary conditions: $\psi(x_0)=\psi_0$ and $\psi(\infty) = 0$, where $x_0$ is the surface position displaced slightly from $x=0$ to avoid the singularity in the non-integer Laplacian operator. 
Here the fifth-order Runge-Kutta method \cite{BVP5c} is employed.
For numerical convergence, we set $x_0$ to be much smaller than the typical ion size $a=1$ nm, and it is confirmed that the numerical error is always less than 0.1$\%$.
Figure 1 shows the comparison between full numerical solution (solid lines) and analytical solution of the linearized PB equation (dashed lines) for different $D$ without the steric effect ($a \to 0, z=1$). 
For low surface potential ($\psi_0=20$ mV), the analytical solutions in all $D$ agree well with the full numerical solutions. 
However, at large surface potential ($\psi_0=50$ mV), the analytical solutions show considerable deviation to the full numerical solutions, especially in small $D$. 
As most practical applications of EDL are operating at low potential, Eq. (11) is a reasonable analytical approximation formula that can be readily employed to extract the value of $D$ from experimental data, thus offering a simplified expression to characterize the degree of spatial complexity in EDL without including the steric effects. 

Figure 2 shows the electrostatic potential profile $\psi$ with an applied surface potential of $\psi_0=100$ mV at different $D$ [Fig. 2(a)] and different $a$ [Fig. 2(b)]. 
For smaller $D <$ 1, the electrostatic potential is decaying slowly, indicating that the screening ability of the counter-ions is reduced. 
This effect is more significant for larger ion size at larger $a$. 
In Fig. 3, the negative ion concentration profile $c^-(x)$ calculated from Eq. (3) is plotted at different $D$ [Fig. 3(a)] and different $a$ [Fig. 3(b)]. Nearer to the charged surface, a saturated layer which has a constant negative ion concentration of $c^-_{sat}$ is formed. 
This saturated concentration $c^-_{sat}$ is independent of $D$ and is always bounded at $1/a^3$. 
In general,  smaller $D$ leads to a wider saturated layer. 
For example, at $c^-(x) = 1$ M, we have about $x$ = 0.36 nm and = 0.54 nm, respectively for $D$ = 1 and 0.1.
The widening of the saturated layer at smaller $D$ again suggests a reduced screening effect at $D <$ 1 in which more counter ions are required to screen out the electrostatic potential generated by the charged electrode.
The effect of different ion sizes on the counter ion concentration profile with a representative value of $D=0.5$ is shown in Fig. 3(b). 
Similar to the $D=1$ case of the BAO model, a higher saturated concentration is reached with smaller ion size, $a$. 

Other than fixed surface potential, we also study the potential and concentration distribution at different $D$ under fixed surface charge density.
By using Eq. (8), the surface charge density is evaluated using the potential gradient profile $\psi^\prime(x)$ obtained from our model. 
Figure 4 shows the electrical potential and negative ion concentration profile plotted with a fixed surface charge density of $\sigma/e = 2$ nm$^{-2}$. 
Similar to the fixed surface potential case, the potential is decaying at a slower speed when $D<1$. This leads to a lower surface potential at smaller $D$ [Fig. 4(a)].
However, the negative ion concentration profile is only slightly affected by $D$ [Fig. 4(b)]. The thickness of the saturated layer with fixed surface charge density is thus not sensitive to $D$ in satisfying the electro-neutrality condition, which is in stark contrast to the case of fixed surface potential depicted in Fig. 3(b).
This finding suggests that, to avoid the sensitivity of spatial complexity (different $D$), one may want to fix the surface charge density instead of surface charge potential in EDL.

In Fig. 5(a), the relationship between the surface negative ion concentration $c_s^-$ to $\sigma/e$ is investigated. 
At low surface charge density where $c_s^-$ does not reach saturation, smaller $D$ leads to a lower value of $c_s^-$. 
As the surface charge density increases, the negative ion concentration saturates at $c_{sat}^-$ ($\approx1/a^3$) at all $D$ . 
We define $\sigma_{sat}$ as the lowest surface charge density required to reach 99\% of $c_{sat}^-$, which is shown in Fig. 5(b) from $D$ = 0.1 to 1. 
The $\sigma_{sat}$ is found to be significantly increased with smaller $D$. 
The higher $\sigma_{sat}$ is a direct consequence of the reduced surface potential with $D<1$, as can be seen in Fig. 4(a).
In Fig. 5(c), the ratio of first layer charge density to surface charge density $\sigma_1/\sigma = zec_s^-a/\sigma$ (a common ratio measured in experimental results \cite{Borukhov,BORUKHOV2000221}) are plotted against $e/\sigma$.
At small $e/\sigma$ (or large $\sigma$) where the surface negative ion concentration is saturated, $\sigma_1$ becomes constant and a linear scaling of $\sigma_1/\sigma$ is observed.
In contrast, the ratio decreases at large $e/\sigma$ (or small $\sigma)$.
The contrasting behaviors of small $e/\sigma$ and large $e/\sigma$ results in a peak of $\sigma_1/\sigma$ ratio, which is plotted as a function of $D$ in Fig. 5(d). 

With X-ray reflectivity experiments, the calculated results above as a function of $D$ could potentially be examined. 
A well-known example is the experiment done by Cuvillier et al. \cite{CUVILLIER199819, cuvillier}, where eicosylamine monolayer were used inside the phosphotungstic acid aqueous solution to form EDL at the air/water interface. 
The surface charge density was controlled by the Langmuir trough lateral pressure, and the electron density profile in the solution was determined through the X-ray reflectivity method. 
As phosphotungstic acid anions have large ion size of $a=1$ nm, its concentration near to the charge surface can reach saturation easily at sufficiently high surface charge density, which is agreeable with our prediction in Fig. 5.
By employing the similar approach in Fig. 5(b), the lowest surface charge density required for the phosphotungstic acid surface anions to reach the saturation ($\sigma_{sat}/e$) can be calculated and plotted as a function of $D$. Using Cuvillier's experiment as an example ($a=1$ nm, $z=3$, $c_0=3\times 10^{-4}$ M), we identify that the values of $D=1,0.7,0.4,0.1$ correspond to $\sigma_{sat}/e=1.05,1.3,1.45,1.6$ nm$^{-2}$ respectively. 
Such empirical fitted $D$ values represent the extend of the spatial inhomogeneity (such as the impurity contained in an electrolyte solution) at the electrode/solution interface, thus simplifying the modelling of the potential and ion distribution in such system. 


\textcolor{blue}{\textbf{Conclusion. }}In conclusion, we have developed a generalized model to account for the spatial complexity in electrical double layer that may be caused by impurities and other inhomogeneity at the electrode/electrolyte interface system. 
The model is based on the fractional calculus approach in which a phenomenal parameter $D$ with a value between 0 and 1 is introduced.
For $D<$ 1, a wider region of saturated ion charge distribution at a fixed surface potential condition is observed, suggesting that the electrostatic potential screening ability of the counter ions is reduced with $D<1$.
This model provides a tool to characterize the degree of spatial complexity in EDL through the extraction of $D$ from the experimental results.
With this known empirical $D$, the model can be used to describe complex EDL structures that are challenging to be characterized directly using the standard PB or BAO methods.
The generalized model presented here offers a new tool to study EDL in a wide range of applications in applied physics and devices \cite{wu2022charge,doi:10.1063/1.3535613}.
With the recent development of three-dimensional AFM imaging technique which provides detailed EDL mapping to atomic level \cite{PhysRevLett.127.196101}, more comprehensive understanding on the relationship between the experiments and our model can be explored in future works.



\section*{ACKNOWLEDGMENTS}
\noindent This work is supported by the USA Office of Naval Research Global (N62909-19-1-2047).
Y.S.A. acknowledge the support of SUTD Start-Up Research Grant (SRG SCI 2021 163). We thank Tingou Liang for conducting preliminary analysis. 

\section*{Author Declarations}
\subsection*{Conflict of Interest}
\noindent The authors declare that there are no conflicts of interest.
\subsection*{Author Contributions}
\noindent C. C. performed the calculation. Y. S. A. conceptualized and initiated the project. Y. S. A. and L. K. A. supervised the project. All authors contributed to the writing of this work.

\section*{Data Availability}
The data that support the findings of this study are available from the corresponding author upon reasonable request.


\begin{thebibliography}{56}%
\makeatletter
\providecommand \@ifxundefined [1]{%
 \@ifx{#1\undefined}
}%
\providecommand \@ifnum [1]{%
 \ifnum #1\expandafter \@firstoftwo
 \else \expandafter \@secondoftwo
 \fi
}%
\providecommand \@ifx [1]{%
 \ifx #1\expandafter \@firstoftwo
 \else \expandafter \@secondoftwo
 \fi
}%
\providecommand \natexlab [1]{#1}%
\providecommand \enquote  [1]{``#1''}%
\providecommand \bibnamefont  [1]{#1}%
\providecommand \bibfnamefont [1]{#1}%
\providecommand \citenamefont [1]{#1}%
\providecommand \href@noop [0]{\@secondoftwo}%
\providecommand \href [0]{\begingroup \@sanitize@url \@href}%
\providecommand \@href[1]{\@@startlink{#1}\@@href}%
\providecommand \@@href[1]{\endgroup#1\@@endlink}%
\providecommand \@sanitize@url [0]{\catcode `\\12\catcode `\$12\catcode
  `\&12\catcode `\#12\catcode `\^12\catcode `\_12\catcode `\%12\relax}%
\providecommand \@@startlink[1]{}%
\providecommand \@@endlink[0]{}%
\providecommand \url  [0]{\begingroup\@sanitize@url \@url }%
\providecommand \@url [1]{\endgroup\@href {#1}{\urlprefix }}%
\providecommand \urlprefix  [0]{URL }%
\providecommand \Eprint [0]{\href }%
\providecommand \doibase [0]{https://doi.org/}%
\providecommand \selectlanguage [0]{\@gobble}%
\providecommand \bibinfo  [0]{\@secondoftwo}%
\providecommand \bibfield  [0]{\@secondoftwo}%
\providecommand \translation [1]{[#1]}%
\providecommand \BibitemOpen [0]{}%
\providecommand \bibitemStop [0]{}%
\providecommand \bibitemNoStop [0]{.\EOS\space}%
\providecommand \EOS [0]{\spacefactor3000\relax}%
\providecommand \BibitemShut  [1]{\csname bibitem#1\endcsname}%
\let\auto@bib@innerbib\@empty
\bibitem [{\citenamefont {Fogolari}\ \emph {et~al.}(1999)\citenamefont
  {Fogolari}, \citenamefont {Zuccato}, \citenamefont {Esposito},\ and\
  \citenamefont {Viglino}}]{bio}%
  \BibitemOpen
  \bibfield  {author} {\bibinfo {author} {\bibfnamefont {F.}~\bibnamefont
  {Fogolari}}, \bibinfo {author} {\bibfnamefont {P.}~\bibnamefont {Zuccato}},
  \bibinfo {author} {\bibfnamefont {G.}~\bibnamefont {Esposito}},\ and\
  \bibinfo {author} {\bibfnamefont {P.}~\bibnamefont {Viglino}},\ }\bibfield
  {title} {\bibinfo {title} {{Biomolecular electrostatics with the linearized
  Poisson-Boltzmann equation}},\ }\href
  {https://doi.org/https://doi.org/10.1016/S0006-3495(99)77173-0} {\bibfield
  {journal} {\bibinfo  {journal} {Biophys. J}\ }\textbf {\bibinfo {volume}
  {76}},\ \bibinfo {pages} {1} (\bibinfo {year} {1999})}\BibitemShut {NoStop}%
\bibitem [{\citenamefont {Blank}(1986)}]{Blank1986ElectricalDL}%
  \BibitemOpen
  \bibfield  {author} {\bibinfo {author} {\bibfnamefont {M.}~\bibnamefont
  {Blank}},\ }\bibfield  {title} {\bibinfo {title} {Electrical double layers in
  biology},\ }in\ \href@noop {} {\emph {\bibinfo {booktitle} {Springer US}}}\
  (\bibinfo {year} {1986})\BibitemShut {NoStop}%
\bibitem [{\citenamefont {Kavanagh}\ \emph {et~al.}(1975)\citenamefont
  {Kavanagh}, \citenamefont {Posner},\ and\ \citenamefont
  {Quirk}}]{DC9755900242}%
  \BibitemOpen
  \bibfield  {author} {\bibinfo {author} {\bibfnamefont {B.~V.}\ \bibnamefont
  {Kavanagh}}, \bibinfo {author} {\bibfnamefont {A.~M.}\ \bibnamefont
  {Posner}},\ and\ \bibinfo {author} {\bibfnamefont {J.~P.}\ \bibnamefont
  {Quirk}},\ }\bibfield  {title} {\bibinfo {title} {Effect of polymer
  adsorption on the properties of the electrical double layer},\ }\href@noop {}
  {\bibfield  {journal} {\bibinfo  {journal} {Faraday Discuss. Chem. Soc.}\
  }\textbf {\bibinfo {volume} {59}},\ \bibinfo {pages} {242} (\bibinfo {year}
  {1975})}\BibitemShut {NoStop}%
\bibitem [{\citenamefont {Nohara}\ \emph {et~al.}(2003)\citenamefont {Nohara},
  \citenamefont {Wada}, \citenamefont {Furukawa}, \citenamefont {Inoue},
  \citenamefont {Morita},\ and\ \citenamefont {Iwakura}}]{NOHARA2003749}%
  \BibitemOpen
  \bibfield  {author} {\bibinfo {author} {\bibfnamefont {S.}~\bibnamefont
  {Nohara}}, \bibinfo {author} {\bibfnamefont {H.}~\bibnamefont {Wada}},
  \bibinfo {author} {\bibfnamefont {N.}~\bibnamefont {Furukawa}}, \bibinfo
  {author} {\bibfnamefont {H.}~\bibnamefont {Inoue}}, \bibinfo {author}
  {\bibfnamefont {M.}~\bibnamefont {Morita}},\ and\ \bibinfo {author}
  {\bibfnamefont {C.}~\bibnamefont {Iwakura}},\ }\bibfield  {title} {\bibinfo
  {title} {Electrochemical characterization of new electric double layer
  capacitor with polymer hydrogel electrolyte},\ }\href
  {https://doi.org/https://doi.org/10.1016/S0013-4686(02)00744-2} {\bibfield
  {journal} {\bibinfo  {journal} {Electrochim. Acta}\ }\textbf {\bibinfo
  {volume} {48}},\ \bibinfo {pages} {749} (\bibinfo {year} {2003})}\BibitemShut
  {NoStop}%
\bibitem [{\citenamefont {Schütter}\ \emph {et~al.}(2019)\citenamefont
  {Schütter}, \citenamefont {Pohlmann},\ and\ \citenamefont
  {Balducci}}]{EDLC}%
  \BibitemOpen
  \bibfield  {author} {\bibinfo {author} {\bibfnamefont {C.}~\bibnamefont
  {Schütter}}, \bibinfo {author} {\bibfnamefont {S.}~\bibnamefont
  {Pohlmann}},\ and\ \bibinfo {author} {\bibfnamefont {A.}~\bibnamefont
  {Balducci}},\ }\bibfield  {title} {\bibinfo {title} {Industrial requirements
  of materials for electrical double layer capacitors: Impact on current and
  future applications},\ }\href@noop {} {\bibfield  {journal} {\bibinfo
  {journal} {Adv. Energy Mater}\ }\textbf {\bibinfo {volume} {9}},\ \bibinfo
  {pages} {1900334} (\bibinfo {year} {2019})}\BibitemShut {NoStop}%
\bibitem [{\citenamefont {Black}\ \emph {et~al.}(2017)\citenamefont {Black},
  \citenamefont {Come}, \citenamefont {Bi}, \citenamefont {Zhu}, \citenamefont
  {Zhao}, \citenamefont {Wong}, \citenamefont {Noh}, \citenamefont {Pudasaini},
  \citenamefont {Zhang}, \citenamefont {Okatan}, \citenamefont {Dai},
  \citenamefont {Kalinin}, \citenamefont {Rack}, \citenamefont {Ward},
  \citenamefont {Feng},\ and\ \citenamefont {Balke}}]{EDLT}%
  \BibitemOpen
  \bibfield  {author} {\bibinfo {author} {\bibfnamefont {J.~M.}\ \bibnamefont
  {Black}}, \bibinfo {author} {\bibfnamefont {J.}~\bibnamefont {Come}},
  \bibinfo {author} {\bibfnamefont {S.}~\bibnamefont {Bi}}, \bibinfo {author}
  {\bibfnamefont {M.}~\bibnamefont {Zhu}}, \bibinfo {author} {\bibfnamefont
  {W.}~\bibnamefont {Zhao}}, \bibinfo {author} {\bibfnamefont {A.~T.}\
  \bibnamefont {Wong}}, \bibinfo {author} {\bibfnamefont {J.~H.}\ \bibnamefont
  {Noh}}, \bibinfo {author} {\bibfnamefont {P.~R.}\ \bibnamefont {Pudasaini}},
  \bibinfo {author} {\bibfnamefont {P.}~\bibnamefont {Zhang}}, \bibinfo
  {author} {\bibfnamefont {M.~B.}\ \bibnamefont {Okatan}}, \bibinfo {author}
  {\bibfnamefont {S.}~\bibnamefont {Dai}}, \bibinfo {author} {\bibfnamefont
  {S.~V.}\ \bibnamefont {Kalinin}}, \bibinfo {author} {\bibfnamefont {P.~D.}\
  \bibnamefont {Rack}}, \bibinfo {author} {\bibfnamefont {T.~Z.}\ \bibnamefont
  {Ward}}, \bibinfo {author} {\bibfnamefont {G.}~\bibnamefont {Feng}},\ and\
  \bibinfo {author} {\bibfnamefont {N.}~\bibnamefont {Balke}},\ }\bibfield
  {title} {\bibinfo {title} {Role of electrical double layer structure in ionic
  liquid gated devices},\ }\href@noop {} {\bibfield  {journal} {\bibinfo
  {journal} {ACS Appl. Mater. Interfaces}\ }\textbf {\bibinfo {volume} {9}},\
  \bibinfo {pages} {40949} (\bibinfo {year} {2017})}\BibitemShut {NoStop}%
\bibitem [{\citenamefont {He}\ \emph {et~al.}(2018)\citenamefont {He},
  \citenamefont {Yang}, \citenamefont {Nie}, \citenamefont {Liu},\ and\
  \citenamefont {Wan}}]{C8TC00530C}%
  \BibitemOpen
  \bibfield  {author} {\bibinfo {author} {\bibfnamefont {Y.}~\bibnamefont
  {He}}, \bibinfo {author} {\bibfnamefont {Y.}~\bibnamefont {Yang}}, \bibinfo
  {author} {\bibfnamefont {S.}~\bibnamefont {Nie}}, \bibinfo {author}
  {\bibfnamefont {R.}~\bibnamefont {Liu}},\ and\ \bibinfo {author}
  {\bibfnamefont {Q.}~\bibnamefont {Wan}},\ }\bibfield  {title} {\bibinfo
  {title} {Electric-double-layer transistors for synaptic devices and
  neuromorphic systems},\ }\href@noop {} {\bibfield  {journal} {\bibinfo
  {journal} {J. Mater. Chem. C}\ }\textbf {\bibinfo {volume} {6}},\ \bibinfo
  {pages} {5336} (\bibinfo {year} {2018})}\BibitemShut {NoStop}%
\bibitem [{\citenamefont {Jiang}\ \emph {et~al.}(2019)\citenamefont {Jiang},
  \citenamefont {Hu}, \citenamefont {Xie}, \citenamefont {Yang}, \citenamefont
  {He}, \citenamefont {Gao},\ and\ \citenamefont {Wan}}]{C8NR07133K}%
  \BibitemOpen
  \bibfield  {author} {\bibinfo {author} {\bibfnamefont {J.}~\bibnamefont
  {Jiang}}, \bibinfo {author} {\bibfnamefont {W.}~\bibnamefont {Hu}}, \bibinfo
  {author} {\bibfnamefont {D.}~\bibnamefont {Xie}}, \bibinfo {author}
  {\bibfnamefont {J.}~\bibnamefont {Yang}}, \bibinfo {author} {\bibfnamefont
  {J.}~\bibnamefont {He}}, \bibinfo {author} {\bibfnamefont {Y.}~\bibnamefont
  {Gao}},\ and\ \bibinfo {author} {\bibfnamefont {Q.}~\bibnamefont {Wan}},\
  }\bibfield  {title} {\bibinfo {title} {{2D electric-double-layer
  phototransistor for photoelectronic and spatiotemporal hybrid neuromorphic
  integration}},\ }\href@noop {} {\bibfield  {journal} {\bibinfo  {journal}
  {Nanoscale}\ }\textbf {\bibinfo {volume} {11}},\ \bibinfo {pages} {1360}
  (\bibinfo {year} {2019})}\BibitemShut {NoStop}%
\bibitem [{\citenamefont {Xie}\ \emph {et~al.}(2018)\citenamefont {Xie},
  \citenamefont {Jiang}, \citenamefont {Hu}, \citenamefont {He}, \citenamefont
  {Yang}, \citenamefont {He}, \citenamefont {Gao},\ and\ \citenamefont
  {Wan}}]{doi:10.1021/acsami.8b07234}%
  \BibitemOpen
  \bibfield  {author} {\bibinfo {author} {\bibfnamefont {D.}~\bibnamefont
  {Xie}}, \bibinfo {author} {\bibfnamefont {J.}~\bibnamefont {Jiang}}, \bibinfo
  {author} {\bibfnamefont {W.}~\bibnamefont {Hu}}, \bibinfo {author}
  {\bibfnamefont {Y.}~\bibnamefont {He}}, \bibinfo {author} {\bibfnamefont
  {J.}~\bibnamefont {Yang}}, \bibinfo {author} {\bibfnamefont {J.}~\bibnamefont
  {He}}, \bibinfo {author} {\bibfnamefont {Y.}~\bibnamefont {Gao}},\ and\
  \bibinfo {author} {\bibfnamefont {Q.}~\bibnamefont {Wan}},\ }\bibfield
  {title} {\bibinfo {title} {{Coplanar multigate MoS$_2$ electric-double-layer
  transistors for neuromorphic visual recognition}},\ }\href
  {https://doi.org/10.1021/acsami.8b07234} {\bibfield  {journal} {\bibinfo
  {journal} {ACS Appl. Mater. Interfaces.}\ }\textbf {\bibinfo {volume} {10}},\
  \bibinfo {pages} {25943} (\bibinfo {year} {2018})}\BibitemShut {NoStop}%
\bibitem [{\citenamefont {Gao}\ \emph {et~al.}(2021)\citenamefont {Gao},
  \citenamefont {Huang}, \citenamefont {Zhang}, \citenamefont {Ji},
  \citenamefont {Zhang}, \citenamefont {Chen}, \citenamefont {Geng},
  \citenamefont {Hu}, \citenamefont {Wang}, \citenamefont {Xiao} \emph
  {et~al.}}]{gao2021artificial}%
  \BibitemOpen
  \bibfield  {author} {\bibinfo {author} {\bibfnamefont {Q.}~\bibnamefont
  {Gao}}, \bibinfo {author} {\bibfnamefont {A.}~\bibnamefont {Huang}}, \bibinfo
  {author} {\bibfnamefont {J.}~\bibnamefont {Zhang}}, \bibinfo {author}
  {\bibfnamefont {Y.}~\bibnamefont {Ji}}, \bibinfo {author} {\bibfnamefont
  {J.}~\bibnamefont {Zhang}}, \bibinfo {author} {\bibfnamefont
  {X.}~\bibnamefont {Chen}}, \bibinfo {author} {\bibfnamefont {X.}~\bibnamefont
  {Geng}}, \bibinfo {author} {\bibfnamefont {Q.}~\bibnamefont {Hu}}, \bibinfo
  {author} {\bibfnamefont {M.}~\bibnamefont {Wang}}, \bibinfo {author}
  {\bibfnamefont {Z.}~\bibnamefont {Xiao}}, \emph {et~al.},\ }\bibfield
  {title} {\bibinfo {title} {Artificial synapses with a sponge-like
  double-layer porous oxide memristor},\ }\href@noop {} {\bibfield  {journal}
  {\bibinfo  {journal} {NPG Asia Mater.}\ }\textbf {\bibinfo {volume} {13}},\
  \bibinfo {pages} {1} (\bibinfo {year} {2021})}\BibitemShut {NoStop}%
\bibitem [{\citenamefont {Gouy}(1910)}]{gouy}%
  \BibitemOpen
  \bibfield  {author} {\bibinfo {author} {\bibfnamefont {M.}~\bibnamefont
  {Gouy}},\ }\bibfield  {title} {\bibinfo {title} {Sur la constitution de la
  charge \'electrique \`a la surface d'un \'electrolyte},\ }\href@noop {}
  {\bibfield  {journal} {\bibinfo  {journal} {J. Phys. Theor. Appl.}\ }\textbf
  {\bibinfo {volume} {9}},\ \bibinfo {pages} {457} (\bibinfo {year}
  {1910})}\BibitemShut {NoStop}%
\bibitem [{\citenamefont {Chapman}(1913)}]{chapman}%
  \BibitemOpen
  \bibfield  {author} {\bibinfo {author} {\bibfnamefont {D.~L.}\ \bibnamefont
  {Chapman}},\ }\bibfield  {title} {\bibinfo {title} {{LI. A contribution to
  the theory of electrocapillarity}},\ }\href
  {https://doi.org/10.1080/14786440408634187} {\bibfield  {journal} {\bibinfo
  {journal} {Lond. Edinb. Dubl. Phil. Mag}\ }\textbf {\bibinfo {volume} {25}},\
  \bibinfo {pages} {475} (\bibinfo {year} {1913})}\BibitemShut {NoStop}%
\bibitem [{\citenamefont {Gray}\ and\ \citenamefont
  {Stiles}(2018)}]{gray2018nonlinear}%
  \BibitemOpen
  \bibfield  {author} {\bibinfo {author} {\bibfnamefont {C.}~\bibnamefont
  {Gray}}\ and\ \bibinfo {author} {\bibfnamefont {P.~J.}\ \bibnamefont
  {Stiles}},\ }\bibfield  {title} {\bibinfo {title} {{Nonlinear electrostatics:
  The Poisson-Boltzmann equation}},\ }\href@noop {} {\bibfield  {journal}
  {\bibinfo  {journal} {Eur. J. Phys.}\ }\textbf {\bibinfo {volume} {39}},\
  \bibinfo {pages} {053002} (\bibinfo {year} {2018})}\BibitemShut {NoStop}%
\bibitem [{\citenamefont {Cuvillier}\ \emph {et~al.}(1997)\citenamefont
  {Cuvillier}, \citenamefont {Bonnier}, \citenamefont {Rondelez}, \citenamefont
  {Paranjape}, \citenamefont {Sastry},\ and\ \citenamefont
  {Ganguly}}]{cuvillier}%
  \BibitemOpen
  \bibfield  {author} {\bibinfo {author} {\bibfnamefont {N.}~\bibnamefont
  {Cuvillier}}, \bibinfo {author} {\bibfnamefont {M.}~\bibnamefont {Bonnier}},
  \bibinfo {author} {\bibfnamefont {F.}~\bibnamefont {Rondelez}}, \bibinfo
  {author} {\bibfnamefont {D.}~\bibnamefont {Paranjape}}, \bibinfo {author}
  {\bibfnamefont {M.}~\bibnamefont {Sastry}},\ and\ \bibinfo {author}
  {\bibfnamefont {P.}~\bibnamefont {Ganguly}},\ }\bibfield  {title} {\bibinfo
  {title} {Adsorption of multivalent ions on charged langmuir monolayers},\
  }in\ \href@noop {} {\emph {\bibinfo {booktitle} {Trends in Colloid and
  Interface Science XI}}},\ \bibinfo {editor} {edited by\ \bibinfo {editor}
  {\bibfnamefont {J.~B.}\ \bibnamefont {Rosenholm}}, \bibinfo {editor}
  {\bibfnamefont {B.}~\bibnamefont {Lindman}},\ and\ \bibinfo {editor}
  {\bibfnamefont {P.}~\bibnamefont {Stenius}}}\ (\bibinfo {year} {1997})\ pp.\
  \bibinfo {pages} {118--125}\BibitemShut {NoStop}%
\bibitem [{\citenamefont {Cuvillier}\ and\ \citenamefont
  {Rondelez}(1998)}]{CUVILLIER199819}%
  \BibitemOpen
  \bibfield  {author} {\bibinfo {author} {\bibfnamefont {N.}~\bibnamefont
  {Cuvillier}}\ and\ \bibinfo {author} {\bibfnamefont {F.}~\bibnamefont
  {Rondelez}},\ }\bibfield  {title} {\bibinfo {title} {{Breakdown of the
  Poisson-Boltzmann description for electrical double layers involving large
  multivalent ions}},\ }\href@noop {} {\bibfield  {journal} {\bibinfo
  {journal} {Thin Solid Films}\ }\textbf {\bibinfo {volume} {327-329}},\
  \bibinfo {pages} {19} (\bibinfo {year} {1998})}\BibitemShut {NoStop}%
\bibitem [{\citenamefont {Brown}\ \emph {et~al.}(2016)\citenamefont {Brown},
  \citenamefont {Abbas}, \citenamefont {Kleibert}, \citenamefont {Green},
  \citenamefont {Goel}, \citenamefont {May},\ and\ \citenamefont
  {Squires}}]{PhysRevX.6.011007}%
  \BibitemOpen
  \bibfield  {author} {\bibinfo {author} {\bibfnamefont {M.~A.}\ \bibnamefont
  {Brown}}, \bibinfo {author} {\bibfnamefont {Z.}~\bibnamefont {Abbas}},
  \bibinfo {author} {\bibfnamefont {A.}~\bibnamefont {Kleibert}}, \bibinfo
  {author} {\bibfnamefont {R.~G.}\ \bibnamefont {Green}}, \bibinfo {author}
  {\bibfnamefont {A.}~\bibnamefont {Goel}}, \bibinfo {author} {\bibfnamefont
  {S.}~\bibnamefont {May}},\ and\ \bibinfo {author} {\bibfnamefont {T.~M.}\
  \bibnamefont {Squires}},\ }\bibfield  {title} {\bibinfo {title}
  {Determination of surface potential and electrical double-layer structure at
  the aqueous electrolyte-nanoparticle interface},\ }\href@noop {} {\bibfield
  {journal} {\bibinfo  {journal} {Phys. Rev. X}\ }\textbf {\bibinfo {volume}
  {6}},\ \bibinfo {pages} {011007} (\bibinfo {year} {2016})}\BibitemShut
  {NoStop}%
\bibitem [{\citenamefont {Butt}\ \emph {et~al.}(2013)\citenamefont {Butt},
  \citenamefont {Graf},\ and\ \citenamefont {Kappl}}]{butt2013physics}%
  \BibitemOpen
  \bibfield  {author} {\bibinfo {author} {\bibfnamefont {H.-J.}\ \bibnamefont
  {Butt}}, \bibinfo {author} {\bibfnamefont {K.}~\bibnamefont {Graf}},\ and\
  \bibinfo {author} {\bibfnamefont {M.}~\bibnamefont {Kappl}},\ }\href@noop {}
  {\emph {\bibinfo {title} {Physics and chemistry of interfaces}}}\ (\bibinfo
  {publisher} {John Wiley \& Sons},\ \bibinfo {year} {2013})\BibitemShut
  {NoStop}%
\bibitem [{\citenamefont {Ji}\ \emph {et~al.}(2020)\citenamefont {Ji},
  \citenamefont {Kim}, \citenamefont {Choi}, \citenamefont {Lee},\ and\
  \citenamefont {Choi}}]{doi:10.1021/acscatal.9b04229}%
  \BibitemOpen
  \bibfield  {author} {\bibinfo {author} {\bibfnamefont {S.~G.}\ \bibnamefont
  {Ji}}, \bibinfo {author} {\bibfnamefont {H.}~\bibnamefont {Kim}}, \bibinfo
  {author} {\bibfnamefont {H.}~\bibnamefont {Choi}}, \bibinfo {author}
  {\bibfnamefont {S.}~\bibnamefont {Lee}},\ and\ \bibinfo {author}
  {\bibfnamefont {C.~H.}\ \bibnamefont {Choi}},\ }\bibfield  {title} {\bibinfo
  {title} {Overestimation of photoelectrochemical hydrogen evolution reactivity
  induced by noble metal impurities dissolved from counter/reference
  electrodes},\ }\href {https://doi.org/10.1021/acscatal.9b04229} {\bibfield
  {journal} {\bibinfo  {journal} {ACS Catal.}\ }\textbf {\bibinfo {volume}
  {10}},\ \bibinfo {pages} {3381} (\bibinfo {year} {2020})}\BibitemShut
  {NoStop}%
\bibitem [{\citenamefont {Lian}\ \emph {et~al.}(2017)\citenamefont {Lian},
  \citenamefont {Liu}, \citenamefont {Liu},\ and\ \citenamefont
  {Wu}}]{doi:10.1021/acs.jpcc.7b04869}%
  \BibitemOpen
  \bibfield  {author} {\bibinfo {author} {\bibfnamefont {C.}~\bibnamefont
  {Lian}}, \bibinfo {author} {\bibfnamefont {K.}~\bibnamefont {Liu}}, \bibinfo
  {author} {\bibfnamefont {H.}~\bibnamefont {Liu}},\ and\ \bibinfo {author}
  {\bibfnamefont {J.}~\bibnamefont {Wu}},\ }\bibfield  {title} {\bibinfo
  {title} {Impurity effects on charging mechanism and energy storage of
  nanoporous supercapacitors},\ }\href
  {https://doi.org/10.1021/acs.jpcc.7b04869} {\bibfield  {journal} {\bibinfo
  {journal} {J. Phys. Chem. C}\ }\textbf {\bibinfo {volume} {121}},\ \bibinfo
  {pages} {14066} (\bibinfo {year} {2017})}\BibitemShut {NoStop}%
\bibitem [{\citenamefont {Borukhov}\ \emph {et~al.}(1997)\citenamefont
  {Borukhov}, \citenamefont {Andelman},\ and\ \citenamefont
  {Orland}}]{Borukhov}%
  \BibitemOpen
  \bibfield  {author} {\bibinfo {author} {\bibfnamefont {I.}~\bibnamefont
  {Borukhov}}, \bibinfo {author} {\bibfnamefont {D.}~\bibnamefont {Andelman}},\
  and\ \bibinfo {author} {\bibfnamefont {H.}~\bibnamefont {Orland}},\
  }\bibfield  {title} {\bibinfo {title} {{Steric effects in electrolytes: A
  modified Poisson-Boltzmann equation}},\ }\href
  {https://doi.org/10.1103/PhysRevLett.79.435} {\bibfield  {journal} {\bibinfo
  {journal} {Phys. Rev. Lett.}\ }\textbf {\bibinfo {volume} {79}},\ \bibinfo
  {pages} {435} (\bibinfo {year} {1997})}\BibitemShut {NoStop}%
\bibitem [{\citenamefont {Borukhov}\ \emph {et~al.}(2000)\citenamefont
  {Borukhov}, \citenamefont {Andelman},\ and\ \citenamefont
  {Orland}}]{BORUKHOV2000221}%
  \BibitemOpen
  \bibfield  {author} {\bibinfo {author} {\bibfnamefont {I.}~\bibnamefont
  {Borukhov}}, \bibinfo {author} {\bibfnamefont {D.}~\bibnamefont {Andelman}},\
  and\ \bibinfo {author} {\bibfnamefont {H.}~\bibnamefont {Orland}},\
  }\bibfield  {title} {\bibinfo {title} {{Adsorption of large ions from an
  electrolyte solution: A modified Poisson–Boltzmann equation}},\ }\href@noop
  {} {\bibfield  {journal} {\bibinfo  {journal} {Electrochim. Acta}\ }\textbf
  {\bibinfo {volume} {46}},\ \bibinfo {pages} {221} (\bibinfo {year}
  {2000})}\BibitemShut {NoStop}%
\bibitem [{\citenamefont {Bohinc}\ and\ \citenamefont {kralj
  iglic}(2001)}]{article}%
  \BibitemOpen
  \bibfield  {author} {\bibinfo {author} {\bibfnamefont {K.}~\bibnamefont
  {Bohinc}}\ and\ \bibinfo {author} {\bibfnamefont {V.}~\bibnamefont {kralj
  iglic}},\ }\bibfield  {title} {\bibinfo {title} {{Thickness of electrical
  double layer. Effect of ion size}},\ }\href
  {https://doi.org/10.1016/S0013-4686(01)00525-4} {\bibfield  {journal}
  {\bibinfo  {journal} {Electrochim. Acta}\ }\textbf {\bibinfo {volume} {46}},\
  \bibinfo {pages} {3033} (\bibinfo {year} {2001})}\BibitemShut {NoStop}%
\bibitem [{\citenamefont {Levine}\ and\ \citenamefont
  {Bell}(1960)}]{doi:10.1021/j100838a019}%
  \BibitemOpen
  \bibfield  {author} {\bibinfo {author} {\bibfnamefont {S.}~\bibnamefont
  {Levine}}\ and\ \bibinfo {author} {\bibfnamefont {G.~M.}\ \bibnamefont
  {Bell}},\ }\bibfield  {title} {\bibinfo {title} {{Theory of a modified
  Poisson-Boltzmann equation. I. The volume effect of hydrated ions}},\
  }\href@noop {} {\bibfield  {journal} {\bibinfo  {journal} {J. Phys. Chem}\
  }\textbf {\bibinfo {volume} {64}},\ \bibinfo {pages} {1188} (\bibinfo {year}
  {1960})}\BibitemShut {NoStop}%
\bibitem [{\citenamefont {Outhwaite}\ and\ \citenamefont
  {Bhuiyan}(1986)}]{doi:10.1063/1.450231}%
  \BibitemOpen
  \bibfield  {author} {\bibinfo {author} {\bibfnamefont {C.~W.}\ \bibnamefont
  {Outhwaite}}\ and\ \bibinfo {author} {\bibfnamefont {L.~B.}\ \bibnamefont
  {Bhuiyan}},\ }\bibfield  {title} {\bibinfo {title} {{A modified
  Poisson–Boltzmann equation in electric double layer theory for a primitive
  model electrolyte with size‐asymmetric ions}},\ }\href
  {https://doi.org/10.1063/1.450231} {\bibfield  {journal} {\bibinfo  {journal}
  {J. Chem. Phys.}\ }\textbf {\bibinfo {volume} {84}},\ \bibinfo {pages} {3461}
  (\bibinfo {year} {1986})}\BibitemShut {NoStop}%
\bibitem [{\citenamefont {Li}\ \emph {et~al.}(2019)\citenamefont {Li},
  \citenamefont {Ying},\ and\ \citenamefont {Xie}}]{li2019analysis}%
  \BibitemOpen
  \bibfield  {author} {\bibinfo {author} {\bibfnamefont {J.}~\bibnamefont
  {Li}}, \bibinfo {author} {\bibfnamefont {J.}~\bibnamefont {Ying}},\ and\
  \bibinfo {author} {\bibfnamefont {D.}~\bibnamefont {Xie}},\ }\bibfield
  {title} {\bibinfo {title} {{On the analysis and application of an ion
  size-modified Poisson-Boltzmann equation}},\ }\href@noop {} {\bibfield
  {journal} {\bibinfo  {journal} {Nonlinear analysis. Real world applications}\
  }\textbf {\bibinfo {volume} {47}},\ \bibinfo {pages} {188} (\bibinfo {year}
  {2019})}\BibitemShut {NoStop}%
\bibitem [{\citenamefont {Potapov}(2007)}]{4368612}%
  \BibitemOpen
  \bibfield  {author} {\bibinfo {author} {\bibfnamefont {A.~A.}\ \bibnamefont
  {Potapov}},\ }\bibfield  {title} {\bibinfo {title} {Theory of fractals and
  fractional dimension in physics and engineering of wave processes},\ }in\
  \href {https://doi.org/10.1109/CRMICO.2007.4368612} {\emph {\bibinfo
  {booktitle} {2007 17th International Crimean Conference - Microwave
  Telecommunication Technology}}}\ (\bibinfo {year} {2007})\ pp.\ \bibinfo
  {pages} {26--27}\BibitemShut {NoStop}%
\bibitem [{\citenamefont {Tarasov}(2015)}]{tarasov}%
  \BibitemOpen
  \bibfield  {author} {\bibinfo {author} {\bibfnamefont {V.}~\bibnamefont
  {Tarasov}},\ }\bibfield  {title} {\bibinfo {title} {Vector calculus in
  non-integer dimensional space and its applications to fractal media},\ }\href
  {https://doi.org/10.1016/j.cnsns.2014.05.025} {\bibfield  {journal} {\bibinfo
   {journal} {Comm. Nonlinear Sci. Numer. Simulat.}\ }\textbf {\bibinfo
  {volume} {20}},\ \bibinfo {pages} {360} (\bibinfo {year} {2015})}\BibitemShut
  {NoStop}%
\bibitem [{\citenamefont {Palmer}\ and\ \citenamefont
  {Stavrinou}(2004)}]{Palmer_2004}%
  \BibitemOpen
  \bibfield  {author} {\bibinfo {author} {\bibfnamefont {C.}~\bibnamefont
  {Palmer}}\ and\ \bibinfo {author} {\bibfnamefont {P.~N.}\ \bibnamefont
  {Stavrinou}},\ }\bibfield  {title} {\bibinfo {title} {Equations of motion in
  a non-integer-dimensional space},\ }\href
  {https://doi.org/10.1088/0305-4470/37/27/009} {\bibfield  {journal} {\bibinfo
   {journal} {J. Phys. A: Math. Theor.}\ }\textbf {\bibinfo {volume} {37}},\
  \bibinfo {pages} {6987} (\bibinfo {year} {2004})}\BibitemShut {NoStop}%
\bibitem [{\citenamefont {Grigorenko}\ and\ \citenamefont
  {Grigorenko}(2003)}]{PhysRevLett.91.034101}%
  \BibitemOpen
  \bibfield  {author} {\bibinfo {author} {\bibfnamefont {I.}~\bibnamefont
  {Grigorenko}}\ and\ \bibinfo {author} {\bibfnamefont {E.}~\bibnamefont
  {Grigorenko}},\ }\bibfield  {title} {\bibinfo {title} {{Chaotic dynamics of
  the fractional Lorenz system}},\ }\href@noop {} {\bibfield  {journal}
  {\bibinfo  {journal} {Phys. Rev. Lett.}\ }\textbf {\bibinfo {volume} {91}},\
  \bibinfo {pages} {034101} (\bibinfo {year} {2003})}\BibitemShut {NoStop}%
\bibitem [{\citenamefont {Paradisi}(2015)}]{paradisi}%
  \BibitemOpen
  \bibfield  {author} {\bibinfo {author} {\bibfnamefont {P.}~\bibnamefont
  {Paradisi}},\ }\bibfield  {title} {\bibinfo {title} {Fractional calculus in
  statistical physics: The case of time fractional diffusion equation},\
  }\href@noop {} {\bibfield  {journal} {\bibinfo  {journal} {Commun. Appl. Ind.
  Math.}\ }\textbf {\bibinfo {volume} {6}} (\bibinfo {year}
  {2015})}\BibitemShut {NoStop}%
\bibitem [{\citenamefont {Tarasov}(2016)}]{TARASOV2016427}%
  \BibitemOpen
  \bibfield  {author} {\bibinfo {author} {\bibfnamefont {V.~E.}\ \bibnamefont
  {Tarasov}},\ }\bibfield  {title} {\bibinfo {title} {Heat transfer in fractal
  materials},\ }\href@noop {} {\bibfield  {journal} {\bibinfo  {journal} {Int.
  J. Heat Mass Transf.}\ }\textbf {\bibinfo {volume} {93}},\ \bibinfo {pages}
  {427} (\bibinfo {year} {2016})}\BibitemShut {NoStop}%
\bibitem [{\citenamefont {Nasrolahpour-Heidari}(2013)}]{nasro}%
  \BibitemOpen
  \bibfield  {author} {\bibinfo {author} {\bibfnamefont {H.}~\bibnamefont
  {Nasrolahpour-Heidari}},\ }\bibfield  {title} {\bibinfo {title} {A note on
  fractional electrodynamics},\ }\href
  {https://doi.org/10.1016/j.cnsns.2013.01.005} {\bibfield  {journal} {\bibinfo
   {journal} {Comm. Nonlinear Sci. Numer. Simulat.}\ }\textbf {\bibinfo
  {volume} {18}},\ \bibinfo {pages} {2589–2593} (\bibinfo {year}
  {2013})}\BibitemShut {NoStop}%
\bibitem [{\citenamefont {Zubair}\ \emph {et~al.}(2012)\citenamefont {Zubair},
  \citenamefont {Mughal},\ and\ \citenamefont {Naqvi}}]{book}%
  \BibitemOpen
  \bibfield  {author} {\bibinfo {author} {\bibfnamefont {M.}~\bibnamefont
  {Zubair}}, \bibinfo {author} {\bibfnamefont {M.~J.}\ \bibnamefont {Mughal}},\
  and\ \bibinfo {author} {\bibfnamefont {Q.~A.}\ \bibnamefont {Naqvi}},\
  }\href@noop {} {\emph {\bibinfo {title} {Electromagnetic fields and waves in
  fractional dimensional space}}}\ (\bibinfo  {publisher} {Springer Science \&
  Business Media},\ \bibinfo {year} {2012})\BibitemShut {NoStop}%
\bibitem [{\citenamefont {He}(1991)}]{PhysRevB.43.2063}%
  \BibitemOpen
  \bibfield  {author} {\bibinfo {author} {\bibfnamefont {X.-F.}\ \bibnamefont
  {He}},\ }\bibfield  {title} {\bibinfo {title} {Excitons in anisotropic
  solids: The model of fractional-dimensional space},\ }\href@noop {}
  {\bibfield  {journal} {\bibinfo  {journal} {Phys. Rev. B}\ }\textbf {\bibinfo
  {volume} {43}},\ \bibinfo {pages} {2063} (\bibinfo {year}
  {1991})}\BibitemShut {NoStop}%
\bibitem [{\citenamefont {Ahmad}\ \emph {et~al.}(2020)\citenamefont {Ahmad},
  \citenamefont {Zubair}, \citenamefont {Jalil}, \citenamefont {Mehmood},
  \citenamefont {Younis}, \citenamefont {Liu}, \citenamefont {Ang},\ and\
  \citenamefont {Ang}}]{excitons-2D}%
  \BibitemOpen
  \bibfield  {author} {\bibinfo {author} {\bibfnamefont {S.}~\bibnamefont
  {Ahmad}}, \bibinfo {author} {\bibfnamefont {M.}~\bibnamefont {Zubair}},
  \bibinfo {author} {\bibfnamefont {O.}~\bibnamefont {Jalil}}, \bibinfo
  {author} {\bibfnamefont {M.~Q.}\ \bibnamefont {Mehmood}}, \bibinfo {author}
  {\bibfnamefont {U.}~\bibnamefont {Younis}}, \bibinfo {author} {\bibfnamefont
  {X.}~\bibnamefont {Liu}}, \bibinfo {author} {\bibfnamefont {K.~W.}\
  \bibnamefont {Ang}},\ and\ \bibinfo {author} {\bibfnamefont {L.~K.}\
  \bibnamefont {Ang}},\ }\bibfield  {title} {\bibinfo {title} {Generalized
  scaling law for exciton binding energy in two-dimensional materials},\
  }\href@noop {} {\bibfield  {journal} {\bibinfo  {journal} {Phys. Rev. Appl.}\
  }\textbf {\bibinfo {volume} {13}},\ \bibinfo {pages} {064062} (\bibinfo
  {year} {2020})}\BibitemShut {NoStop}%
\bibitem [{\citenamefont {Mondol}\ \emph {et~al.}(2018)\citenamefont {Mondol},
  \citenamefont {Gupta}, \citenamefont {Das},\ and\ \citenamefont
  {Dutta}}]{doi:10.1063/1.4998236}%
  \BibitemOpen
  \bibfield  {author} {\bibinfo {author} {\bibfnamefont {A.}~\bibnamefont
  {Mondol}}, \bibinfo {author} {\bibfnamefont {R.}~\bibnamefont {Gupta}},
  \bibinfo {author} {\bibfnamefont {S.}~\bibnamefont {Das}},\ and\ \bibinfo
  {author} {\bibfnamefont {T.}~\bibnamefont {Dutta}},\ }\bibfield  {title}
  {\bibinfo {title} {{An insight into Newton's cooling law using fractional
  calculus}},\ }\href@noop {} {\bibfield  {journal} {\bibinfo  {journal} {J.
  Appl. Phys.}\ }\textbf {\bibinfo {volume} {123}},\ \bibinfo {pages} {064901}
  (\bibinfo {year} {2018})}\BibitemShut {NoStop}%
\bibitem [{\citenamefont {Zubair}\ \emph
  {et~al.}(2018{\natexlab{a}})\citenamefont {Zubair}, \citenamefont {Ang},\
  and\ \citenamefont {Ang}}]{8388713}%
  \BibitemOpen
  \bibfield  {author} {\bibinfo {author} {\bibfnamefont {M.}~\bibnamefont
  {Zubair}}, \bibinfo {author} {\bibfnamefont {Y.~S.}\ \bibnamefont {Ang}},\
  and\ \bibinfo {author} {\bibfnamefont {L.~K.}\ \bibnamefont {Ang}},\
  }\bibfield  {title} {\bibinfo {title} {Thickness dependence of
  space-charge-limited current in spatially disordered organic
  semiconductors},\ }\href {https://doi.org/10.1109/TED.2018.2841920}
  {\bibfield  {journal} {\bibinfo  {journal} {IEEE Trans. Electron Devices}\
  }\textbf {\bibinfo {volume} {65}},\ \bibinfo {pages} {3421} (\bibinfo {year}
  {2018}{\natexlab{a}})}\BibitemShut {NoStop}%
\bibitem [{\citenamefont {Zubair}\ \emph
  {et~al.}(2018{\natexlab{b}})\citenamefont {Zubair}, \citenamefont {Ang},\
  and\ \citenamefont {Ang}}]{fractionalFN}%
  \BibitemOpen
  \bibfield  {author} {\bibinfo {author} {\bibfnamefont {M.}~\bibnamefont
  {Zubair}}, \bibinfo {author} {\bibfnamefont {Y.~S.}\ \bibnamefont {Ang}},\
  and\ \bibinfo {author} {\bibfnamefont {L.~K.}\ \bibnamefont {Ang}},\
  }\bibfield  {title} {\bibinfo {title} {Fractional fowler--nordheim law for
  field emission from rough surface with nonparabolic energy dispersion},\
  }\href@noop {} {\bibfield  {journal} {\bibinfo  {journal} {IEEE Trans.
  Electron Devices}\ }\textbf {\bibinfo {volume} {65}},\ \bibinfo {pages}
  {2089} (\bibinfo {year} {2018}{\natexlab{b}})}\BibitemShut {NoStop}%
\bibitem [{\citenamefont {Zubair}\ and\ \citenamefont
  {Ang}(2016)}]{doi:10.1063/1.4958944}%
  \BibitemOpen
  \bibfield  {author} {\bibinfo {author} {\bibfnamefont {M.}~\bibnamefont
  {Zubair}}\ and\ \bibinfo {author} {\bibfnamefont {L.~K.}\ \bibnamefont
  {Ang}},\ }\bibfield  {title} {\bibinfo {title} {{Fractional-dimensional
  Child-Langmuir law for a rough cathode}},\ }\href
  {https://doi.org/10.1063/1.4958944} {\bibfield  {journal} {\bibinfo
  {journal} {Phys. Plasmas}\ }\textbf {\bibinfo {volume} {23}},\ \bibinfo
  {pages} {072118} (\bibinfo {year} {2016})}\BibitemShut {NoStop}%
\bibitem [{\citenamefont {Zubair}\ \emph
  {et~al.}(2018{\natexlab{c}})\citenamefont {Zubair}, \citenamefont {Ang},
  \citenamefont {Ooi},\ and\ \citenamefont {Ang}}]{doi:10.1063/1.5039811}%
  \BibitemOpen
  \bibfield  {author} {\bibinfo {author} {\bibfnamefont {M.}~\bibnamefont
  {Zubair}}, \bibinfo {author} {\bibfnamefont {Y.~S.}\ \bibnamefont {Ang}},
  \bibinfo {author} {\bibfnamefont {K.~J.~A.}\ \bibnamefont {Ooi}},\ and\
  \bibinfo {author} {\bibfnamefont {L.~K.}\ \bibnamefont {Ang}},\ }\bibfield
  {title} {\bibinfo {title} {{Fractional Fresnel coefficients for optical
  absorption in femtosecond laser-induced rough metal surfaces}},\ }\href@noop
  {} {\bibfield  {journal} {\bibinfo  {journal} {J. Appl. Phys.}\ }\textbf
  {\bibinfo {volume} {124}},\ \bibinfo {pages} {163101} (\bibinfo {year}
  {2018}{\natexlab{c}})}\BibitemShut {NoStop}%
\bibitem [{\citenamefont {Zubair}\ \emph {et~al.}(2011)\citenamefont {Zubair},
  \citenamefont {Mughal},\ and\ \citenamefont {Naqvi}}]{zubair2011exact}%
  \BibitemOpen
  \bibfield  {author} {\bibinfo {author} {\bibfnamefont {M.}~\bibnamefont
  {Zubair}}, \bibinfo {author} {\bibfnamefont {M.~J.}\ \bibnamefont {Mughal}},\
  and\ \bibinfo {author} {\bibfnamefont {Q.}~\bibnamefont {Naqvi}},\ }\bibfield
   {title} {\bibinfo {title} {An exact solution of the cylindrical wave
  equation for electromagnetic field in fractional dimensional space},\
  }\href@noop {} {\bibfield  {journal} {\bibinfo  {journal} {Prog. Electromagn.
  Res.}\ }\textbf {\bibinfo {volume} {114}},\ \bibinfo {pages} {443} (\bibinfo
  {year} {2011})}\BibitemShut {NoStop}%
\bibitem [{\citenamefont {Ehsan}\ \emph {et~al.}(2020)\citenamefont {Ehsan},
  \citenamefont {Mehmood}, \citenamefont {Ang}, \citenamefont {Ang},\ and\
  \citenamefont {Zubair}}]{Ehsan2020SpaceFractionalBB}%
  \BibitemOpen
  \bibfield  {author} {\bibinfo {author} {\bibfnamefont {A.}~\bibnamefont
  {Ehsan}}, \bibinfo {author} {\bibfnamefont {M.~Q.}\ \bibnamefont {Mehmood}},
  \bibinfo {author} {\bibfnamefont {Y.~S.}\ \bibnamefont {Ang}}, \bibinfo
  {author} {\bibfnamefont {L.~K.}\ \bibnamefont {Ang}},\ and\ \bibinfo {author}
  {\bibfnamefont {M.}~\bibnamefont {Zubair}},\ }\bibfield  {title} {\bibinfo
  {title} {{Space-fractional Bessel beams with self-healing and
  diffraction-free propagation characteristics}},\ }in\ \href
  {https://doi.org/10.23919/EuCAP48036.2020.9135195} {\emph {\bibinfo
  {booktitle} {2020 14th European Conference on Antennas and Propagation
  (EuCAP)}}}\ (\bibinfo {year} {2020})\ pp.\ \bibinfo {pages}
  {1--5}\BibitemShut {NoStop}%
\bibitem [{\citenamefont {Naqvi}\ \emph {et~al.}(2010)\citenamefont {Naqvi},
  \citenamefont {Hussain},\ and\ \citenamefont {Naqvi}}]{naqvi2010waves}%
  \BibitemOpen
  \bibfield  {author} {\bibinfo {author} {\bibfnamefont {A.}~\bibnamefont
  {Naqvi}}, \bibinfo {author} {\bibfnamefont {A.}~\bibnamefont {Hussain}},\
  and\ \bibinfo {author} {\bibfnamefont {Q.}~\bibnamefont {Naqvi}},\ }\bibfield
   {title} {\bibinfo {title} {Waves in fractional dual planar waveguides
  containing chiral nihility metamaterial},\ }\href@noop {} {\bibfield
  {journal} {\bibinfo  {journal} {J. Electromagn. Waves Appl.}\ }\textbf
  {\bibinfo {volume} {24}},\ \bibinfo {pages} {1575} (\bibinfo {year}
  {2010})}\BibitemShut {NoStop}%
\bibitem [{\citenamefont {Nisar}\ and\ \citenamefont
  {Naqvi}(2018)}]{nisar2018cloaking}%
  \BibitemOpen
  \bibfield  {author} {\bibinfo {author} {\bibfnamefont {M.}~\bibnamefont
  {Nisar}}\ and\ \bibinfo {author} {\bibfnamefont {Q.~A.}\ \bibnamefont
  {Naqvi}},\ }\bibfield  {title} {\bibinfo {title} {Cloaking and magnifying
  using radial anisotropy in non-integer dimensional space},\ }\href@noop {}
  {\bibfield  {journal} {\bibinfo  {journal} {Phys. Lett. A}\ }\textbf
  {\bibinfo {volume} {382}},\ \bibinfo {pages} {2055} (\bibinfo {year}
  {2018})}\BibitemShut {NoStop}%
\bibitem [{\citenamefont {Tuan}\ \emph {et~al.}(2020)\citenamefont {Tuan},
  \citenamefont {Mohammadi},\ and\ \citenamefont
  {Rezapour}}]{tuan2020mathematical}%
  \BibitemOpen
  \bibfield  {author} {\bibinfo {author} {\bibfnamefont {N.~H.}\ \bibnamefont
  {Tuan}}, \bibinfo {author} {\bibfnamefont {H.}~\bibnamefont {Mohammadi}},\
  and\ \bibinfo {author} {\bibfnamefont {S.}~\bibnamefont {Rezapour}},\
  }\bibfield  {title} {\bibinfo {title} {{A mathematical model for COVID-19
  transmission by using the Caputo fractional derivative}},\ }\href@noop {}
  {\bibfield  {journal} {\bibinfo  {journal} {Chaos Solit. Fractals}\ }\textbf
  {\bibinfo {volume} {140}},\ \bibinfo {pages} {110107} (\bibinfo {year}
  {2020})}\BibitemShut {NoStop}%
\bibitem [{\citenamefont {Rajagopal}\ \emph {et~al.}(2020)\citenamefont
  {Rajagopal}, \citenamefont {Hasanzadeh}, \citenamefont {Parastesh},
  \citenamefont {Hamarash}, \citenamefont {Jafari},\ and\ \citenamefont
  {Hussain}}]{Rajagopal2020AFM}%
  \BibitemOpen
  \bibfield  {author} {\bibinfo {author} {\bibfnamefont {K.}~\bibnamefont
  {Rajagopal}}, \bibinfo {author} {\bibfnamefont {N.}~\bibnamefont
  {Hasanzadeh}}, \bibinfo {author} {\bibfnamefont {F.}~\bibnamefont
  {Parastesh}}, \bibinfo {author} {\bibfnamefont {I.~I.}\ \bibnamefont
  {Hamarash}}, \bibinfo {author} {\bibfnamefont {S.}~\bibnamefont {Jafari}},\
  and\ \bibinfo {author} {\bibfnamefont {I.}~\bibnamefont {Hussain}},\
  }\bibfield  {title} {\bibinfo {title} {A fractional-order model for the novel
  coronavirus (covid-19) outbreak},\ }\href@noop {} {\bibfield  {journal}
  {\bibinfo  {journal} {Nonlinear Dyn.}\ ,\ \bibinfo {pages} {1 }} (\bibinfo
  {year} {2020})}\BibitemShut {NoStop}%
\bibitem [{\citenamefont {Yuan}\ \emph {et~al.}(2014)\citenamefont {Yuan},
  \citenamefont {Fu},\ and\ \citenamefont {Liu}}]{Climate}%
  \BibitemOpen
  \bibfield  {author} {\bibinfo {author} {\bibfnamefont {N.}~\bibnamefont
  {Yuan}}, \bibinfo {author} {\bibfnamefont {Z.}~\bibnamefont {Fu}},\ and\
  \bibinfo {author} {\bibfnamefont {S.}~\bibnamefont {Liu}},\ }\bibfield
  {title} {\bibinfo {title} {Extracting climate memory using fractional
  integrated statistical model: A new perspective on climate prediction},\
  }\href {https://doi.org/10.1038/srep06577} {\bibfield  {journal} {\bibinfo
  {journal} {Sci. Rep.}\ }\textbf {\bibinfo {volume} {4}},\ \bibinfo {pages}
  {6577} (\bibinfo {year} {2014})}\BibitemShut {NoStop}%
\bibitem [{\citenamefont {Tarasova}\ and\ \citenamefont
  {Tarasov}(2016{\natexlab{a}})}]{tarasova2016elasticity}%
  \BibitemOpen
  \bibfield  {author} {\bibinfo {author} {\bibfnamefont {V.~V.}\ \bibnamefont
  {Tarasova}}\ and\ \bibinfo {author} {\bibfnamefont {V.~E.}\ \bibnamefont
  {Tarasov}},\ }\bibfield  {title} {\bibinfo {title} {Elasticity for economic
  processes with memory: Fractional differential calculus approach},\
  }\href@noop {} {\bibfield  {journal} {\bibinfo  {journal} {Fract. Differ.
  Calc}\ }\textbf {\bibinfo {volume} {6}},\ \bibinfo {pages} {219} (\bibinfo
  {year} {2016}{\natexlab{a}})}\BibitemShut {NoStop}%
\bibitem [{\citenamefont {Tarasov}(2019)}]{math7060509}%
  \BibitemOpen
  \bibfield  {author} {\bibinfo {author} {\bibfnamefont {V.~E.}\ \bibnamefont
  {Tarasov}},\ }\bibfield  {title} {\bibinfo {title} {On history of
  mathematical economics: Application of fractional calculus},\ }\href@noop {}
  {\bibfield  {journal} {\bibinfo  {journal} {Mathematics}\ }\textbf {\bibinfo
  {volume} {7}},\ \bibinfo {pages} {509} (\bibinfo {year} {2019})}\BibitemShut
  {NoStop}%
\bibitem [{\citenamefont {Tarasova}\ and\ \citenamefont
  {Tarasov}(2016{\natexlab{b}})}]{Tarasov2016memory}%
  \BibitemOpen
  \bibfield  {author} {\bibinfo {author} {\bibfnamefont {V.}~\bibnamefont
  {Tarasova}}\ and\ \bibinfo {author} {\bibfnamefont {V.}~\bibnamefont
  {Tarasov}},\ }\bibfield  {title} {\bibinfo {title} {Fractional dynamics of
  natural growth and memory effect in economics},\ }\href@noop {} {\bibfield
  {journal} {\bibinfo  {journal} {Eur. Res.}\ }\textbf {\bibinfo {volume} {23}}
  (\bibinfo {year} {2016}{\natexlab{b}})}\BibitemShut {NoStop}%
\bibitem [{\citenamefont {Kumar}\ \emph {et~al.}(2021)\citenamefont {Kumar},
  \citenamefont {Erturk},\ and\ \citenamefont
  {Murillo-Arcila}}]{KUMAR2021111091}%
  \BibitemOpen
  \bibfield  {author} {\bibinfo {author} {\bibfnamefont {P.}~\bibnamefont
  {Kumar}}, \bibinfo {author} {\bibfnamefont {V.~S.}\ \bibnamefont {Erturk}},\
  and\ \bibinfo {author} {\bibfnamefont {M.}~\bibnamefont {Murillo-Arcila}},\
  }\bibfield  {title} {\bibinfo {title} {{A complex fractional mathematical
  modeling for the love story of Layla and Majnun}},\ }\href@noop {} {\bibfield
   {journal} {\bibinfo  {journal} {Chaos Solit. Fractals}\ }\textbf {\bibinfo
  {volume} {150}},\ \bibinfo {pages} {111091} (\bibinfo {year}
  {2021})}\BibitemShut {NoStop}%
\bibitem [{\citenamefont {Stillinger}(1977)}]{stillinger}%
  \BibitemOpen
  \bibfield  {author} {\bibinfo {author} {\bibfnamefont {F.}~\bibnamefont
  {Stillinger}},\ }\bibfield  {title} {\bibinfo {title} {Axiomatic basis for
  spaces with noninteger dimension},\ }\href@noop {} {\bibfield  {journal}
  {\bibinfo  {journal} {J. Math. Phys}\ }\textbf {\bibinfo {volume} {18}},\
  \bibinfo {pages} {1224} (\bibinfo {year} {1977})}\BibitemShut {NoStop}%
\bibitem [{\citenamefont {Kierzenka}\ and\ \citenamefont
  {Shampine}(2008)}]{BVP5c}%
  \BibitemOpen
  \bibfield  {author} {\bibinfo {author} {\bibfnamefont {J.}~\bibnamefont
  {Kierzenka}}\ and\ \bibinfo {author} {\bibfnamefont {L.}~\bibnamefont
  {Shampine}},\ }\bibfield  {title} {\bibinfo {title} {{A BVP solver that
  controls residual and error}},\ }\href@noop {} {\bibfield  {journal}
  {\bibinfo  {journal} {J. Numer. Anal. Ind. Appl. Math.}\ }\textbf {\bibinfo
  {volume} {3}},\ \bibinfo {pages} {27} (\bibinfo {year} {2008})}\BibitemShut
  {NoStop}%
\bibitem [{\citenamefont {Wu}\ and\ \citenamefont
  {Senevirathna}(2022)}]{wu2022charge}%
  \BibitemOpen
  \bibfield  {author} {\bibinfo {author} {\bibfnamefont {P.}~\bibnamefont
  {Wu}}\ and\ \bibinfo {author} {\bibfnamefont {H.~L.}\ \bibnamefont
  {Senevirathna}},\ }\bibfield  {title} {\bibinfo {title} {A charge transport
  prediction method for metal electrode-aqueous electrolyte interface to guide
  the design of metal batteries},\ }\href@noop {} {\bibfield  {journal}
  {\bibinfo  {journal} {Electrochim. Acta}\ ,\ \bibinfo {pages} {140051}}
  (\bibinfo {year} {2022})}\BibitemShut {NoStop}%
\bibitem [{\citenamefont {Yuan}\ \emph {et~al.}(2011)\citenamefont {Yuan},
  \citenamefont {Toh}, \citenamefont {Morimoto}, \citenamefont {Tan},
  \citenamefont {Wei}, \citenamefont {Shimotani}, \citenamefont {Kloc},\ and\
  \citenamefont {Iwasa}}]{doi:10.1063/1.3535613}%
  \BibitemOpen
  \bibfield  {author} {\bibinfo {author} {\bibfnamefont {H.~T.}\ \bibnamefont
  {Yuan}}, \bibinfo {author} {\bibfnamefont {M.}~\bibnamefont {Toh}}, \bibinfo
  {author} {\bibfnamefont {K.}~\bibnamefont {Morimoto}}, \bibinfo {author}
  {\bibfnamefont {W.}~\bibnamefont {Tan}}, \bibinfo {author} {\bibfnamefont
  {F.}~\bibnamefont {Wei}}, \bibinfo {author} {\bibfnamefont {H.}~\bibnamefont
  {Shimotani}}, \bibinfo {author} {\bibfnamefont {C.}~\bibnamefont {Kloc}},\
  and\ \bibinfo {author} {\bibfnamefont {Y.}~\bibnamefont {Iwasa}},\ }\bibfield
   {title} {\bibinfo {title} {{Liquid-gated electric-double-layer transistor on
  layered metal dichalcogenide, SnS2}},\ }\href@noop {} {\bibfield  {journal}
  {\bibinfo  {journal} {Appl. Phys}\ }\textbf {\bibinfo {volume} {98}},\
  \bibinfo {pages} {012102} (\bibinfo {year} {2011})}\BibitemShut {NoStop}%
\bibitem [{\citenamefont {Benaglia}\ \emph {et~al.}(2021)\citenamefont
  {Benaglia}, \citenamefont {Uhlig}, \citenamefont {Hern\'andez-Mu\~noz},
  \citenamefont {Chac\'on}, \citenamefont {Tarazona},\ and\ \citenamefont
  {Garcia}}]{PhysRevLett.127.196101}%
  \BibitemOpen
  \bibfield  {author} {\bibinfo {author} {\bibfnamefont {S.}~\bibnamefont
  {Benaglia}}, \bibinfo {author} {\bibfnamefont {M.~R.}\ \bibnamefont {Uhlig}},
  \bibinfo {author} {\bibfnamefont {J.}~\bibnamefont {Hern\'andez-Mu\~noz}},
  \bibinfo {author} {\bibfnamefont {E.}~\bibnamefont {Chac\'on}}, \bibinfo
  {author} {\bibfnamefont {P.}~\bibnamefont {Tarazona}},\ and\ \bibinfo
  {author} {\bibfnamefont {R.}~\bibnamefont {Garcia}},\ }\bibfield  {title}
  {\bibinfo {title} {{Tip charge dependence of three-dimensional AFM mapping of
  concentrated ionic solutions}},\ }\href@noop {} {\bibfield  {journal}
  {\bibinfo  {journal} {Phys. Rev. Lett.}\ }\textbf {\bibinfo {volume} {127}},\
  \bibinfo {pages} {196101} (\bibinfo {year} {2021})}\BibitemShut {NoStop}%
\end{thebibliography}

%

\end{document}